\documentclass[cup6a]{cupbook}
\usepackage{graphicx}
\title[Dark energy, gravitation and the Copernican principle]
      {Dark energy, gravitation and the Copernican principle}
\author{Jean-Philippe Uzan}
\date{15th August 2008}

\begin{document}

\pagenumbering{roman}
\cleardoublepage
\pagenumbering{arabic}

\def\dd{{\rm d}}
\def\Box{(\nabla_\mu\nabla^\mu)}
\def\eff{{\rm eff}}
\def\ppn{{\rm PPN}}
\def\cav{\mathrm{cav}}
\def\url#1{[{\tt #1}]}
\def\de{{\rm de}}
\def\mat{{\rm m}}
\def\etal{et al.{$\,$}}

\chapter{Dark energy, gravitation and the Copernican principle}
\author{Jean-Philippe Uzan\\
             Institut d'Astrophysique de Paris,
             Universit\'e Pierre~\&~Marie Curie - Paris VI,
             CNRS-UMR 7095, 98 bis, Bd Arago, 75014 Paris, France.}

\begin{center}
             Jean-Philippe Uzan\\
             Institut d'Astrophysique de Paris, CNRS-UMR 7095,\\
             Universit\'e Pierre~\&~Marie Curie - Paris~VI,\\
              98 bis, Bd Arago, 75014 Paris, France.\\
             \vskip1cm
             (15th August 2008)
             \vskip1cm
             {To appear in {\it Dark Energy: Observational and Theoretical
Approaches},\\ Ed. P. Ruiz-Lapuente (Cambridge University Press, 2010).}
\end{center}

\section{Cosmological models and their hypotheses}

\subsection{Introduction}

The progresses of physical cosmology during the past ten years
have led to a ``standard'' cosmological model in agreement with
all available data. Its parameters are measured with increasing
precision but it requires the introduction of a dark sector,
including both dark matter and dark energy, attracting the
attention of both observers and theoreticians.

Among all the observational conclusions, the existence of a recent
acceleration phase of the cosmic expansion is more and more
robust. The quest for the understanding of its physical origin is
however just starting~(Peebles and Ratra, 2003; Peter and Uzan,
2005; Copeland \etal, 2006; Uzan, 2007). Models and speculations
are flourishing and we may wonder to which extent the observations
of our local Universe may reveal the physical nature of the dark
energy. In particular, there exist limitations to this quest
intrinsic to cosmology, related to the fact that most observations
are located on our past light-cone~(Ellis, 1975), and to finite
volume effects~(Bernardeau and Uzan, 2004) that can make many
physically acceptable possibilities undistinguishable in practice.

This text aims at discussing the relations between the cosmic
acceleration and the theory of gravitation and more generally with
the hypotheses underlying the construction of our cosmological
model, such as the validity of general relativity on astrophysical
scales and the Copernican principle. We hope to illustrate that
cosmological data have now the potential of testing these
hypotheses, which go beyond the measurements of its parameters.

\subsection{Cosmology, physics and astronomy}

Cosmology seats at the cross-road between theoretical physics and
astronomy.

Theoretical physics, based on physical laws, tries to describe the
fundamental components of nature and their interactions. These
laws can be probed locally by experiments. These laws need to be
extrapolated to construct cosmological models. Hence any new idea
or discovery concerning these laws can naturally call for an
extension of our cosmological model (e.g. introducing massive
neutrinos in cosmology is now mandatory).

Astronomy confronts us with phenomena that we have to understand
and explain consistently. This often requires the introduction of
hypotheses beyond those of the physical theories (\S~\ref{sec113})
in order to ``{\em save the phenomena}''~(Duhem, 1908), as is
actually the case with the dark sector of our cosmological model.
Needless to remind that even if a cosmological model is in
agreement with all observations, whatever their accuracy, it does
not prove that it is the ``correct'' model of the Universe, in the
sense that it is the correct cosmological extrapolation and
solution of the local physical laws.

Dark energy confronts us with a compatibility problem since, in
order to ``save the phenomena'' of the observations, we have to
include new ingredients (constant, matter fields or interactions)
beyond those of our established physical theories. However the
required value for the simplest dark energy model, i.e. the
cosmological constant, is more than 60 order of magnitude smaller
to what is expected from theoretical grounds (\S~\ref{sec116}).
This tension between what is required by astronomy and what is
expected from physics reminds us of the twenty centuries long
debate between Aristotelians and Ptolemeans~(Duhem, 1913), that
was resolved not only by the Copernican model but more important
by a better understanding of the physics since Newton gravity was
compatible only with one of these three models that, at the time,
could not be distinguished observationally.

\subsection{hypotheses of our cosmological model}\label{sec113}

The construction of any cosmological model relies on 4 main
hypotheses,
\begin{itemize}
  \item[(H1)] a theory of gravity,
  \item[(H2)] a description of the matter contained in the Universe and their
  non-gravitational interactions,
  \item[(H3)] symmetry hypotheses, and
  \item[(H4)] an hypothesis on the global structure, i.e. the topology,
  of the Universe.
\end{itemize}
These hypotheses are not on the same footing since H1 and H2 refer
to the physical theories. These two hypotheses are however not
sufficient to solve the field equations and we must make an
assumption on the symmetries (H3) of the solutions describing our
Universe on large scales while H4 is an assumption on some global
properties of these cosmological solutions, with same local
geometry.

Our reference cosmological model is the $\Lambda$CDM model. It
assumes that gravity is described by general relativity (H1), that
the Universe contains the fields of the standard model of particle
physics plus some dark matter and a cosmological constant, the
latter two having no physical explanation at the moment. Note that
in the cosmological context this involves an extra-assumption
since what will be required by the Einstein equations is the
effective stress-energy tensor averaged on large scales. It thus
implicitly refers to a, usually not explicited, averaging
procedure~(Ellis and Buchert, 2005). It also deeply involves the
Copernican principle as a symmetry hypotheses (H3), without which
the Einstein equations usually can not been solved, and usually
assumes that the spatial sections are simply connected (H4). H2
and H3 imply that the description of standard matter reduces to a
mixture of a pressureless and a radiation perfect fluids.

\subsection{Copernican principle}

The {\it cosmological principle} supposes that the Universe is
spatially isotropic and homogeneous. In particular, this implies
that there exists a privileged class of observers, called
fundamental observers, who all see an isotropic universe around
them. It implies the existence of a cosmic time and states that
all the properties of the universe are the same everywhere at the
same cosmic time. It is supposed to hold for the smooth-out
structure of the Universe on large scales. Indeed, this principle
has to be applied in a statistical sense since there exist
structures in the universe.

We can distinguish it from the {\it Copernican principle} which
merely states that we do not live in a special place (the center)
of the Universe. As long as isotropy around the observer holds,
the principle actually leads to the same conclusion than the
cosmological principle.

The cosmological principle makes definite predictions about all
unobservable regions beyond the observable universe. It completely
determines the entire structure of the Universe, even for regions
that cannot be observed. From this point of view, this hypothesis,
which cannot be tested, is very strong. On the other hand it leads
to a complete model of the universe. The Copernican principle has
more modest consequences and leads to the same conclusions but
only for the observable universe where isotropy has been verified.
It does not make any prediction on the structure of the Universe
for unobserved regions (in particular, space could be homogeneous
and non isotropic on scales larger than the observable Universe).
We refer to Bondi (1960), North (1965) and Ellis (1975) for
further discussions on the definition of these two principles.

We emphasized that, as shall be discussed in the next section, our
reference cosmological model includes a primordial phase of
inflation in order to explain the origin of the large scale
structures of the Universe. Inflation gives a theoretical
prejudice in favor of the Copernican principle since it predicts
that all classical (i.e. non-quantum) inhomogeneities (curvature,
shear, \ldots) have been washed-out during this phase. If it is
sufficiently long, we expect the principle to hold on scales much
larger than those of the observable universe, hence backing-up the
cosmological principle, since unobservable regions today arise
from the same causal process that affected the conditions in our
local Universe. While the standard predictions of inflation are in
agreement with all astronomical data, we should not forget it is
only a theoretical argument on which we shall come back in the
case we find observable evidences against isotropy~(Pereira \etal,
2007; Pitrou \etal, 2008), curvature~(Uzan \etal, 2003) and
homogeneity (e.g. such as a spatial topology of the Universe).

This principles leads to a Robertson-Walker (RW) geometry  with
metric
\begin{equation}\label{flrw}
 \dd s^2 = -\dd t^2 + a^2(t)\gamma_{ij}\dd x^i\dd x^j,
\end{equation}
where $t$ is the cosmic time and $\gamma_{ij}$ is the spatial
metric on the constant time hypersurfaces, which are homogeneous
and isotropic, and thus of constant curvature. It follows that the
metric is reduced to a single function of time, the scale factor.
This implies that there is a one-to-one mapping between the cosmic
time and the redshift
\begin{equation}\label{z.flrw}
 1+z =\frac{a_0}{a(t)},
\end{equation}
if the expansion is monotonous.

\subsection{$\Lambda$CDM reference model}

The dynamics of the scale factor can be determined from the
Einstein equations which reduce for the metric~(\ref{flrw}) to the
Friedmann equations
\begin{eqnarray}
 H^2 &=& \frac{8\pi G}{3}\rho - \frac{K}{a^2} + \frac{\Lambda}{3},\\
 \frac{\ddot a}{a} &=& -\frac{4\pi G}{3}(\rho+3P)+
 \frac{\Lambda}{3}\label{fried2}.
\end{eqnarray}
$H\equiv\dot a/a$ is the Hubble function and $K=0,\pm1$ is the
curvature of the spatial sections. $G$ and $\Lambda$ are the
Newton and cosmological constants. $\rho$ and $P$ are respectively
the energy density and pressure of the cosmic fluids and are
related by
$$
 \dot\rho+3H(\rho+P)=0.
$$
Defining the dimensionless density parameters as
\begin{equation}\label{3.omega}
 \Omega=\frac{8\pi G\rho}{3H^2}\ ,\quad
 \Omega_\Lambda=\frac{\Lambda}{3H^2}\ ,\quad
 \Omega_K=-\frac{K}{H^2a^2},
\end{equation}
respectively for the matter, the cosmological constant and the
curvature, the first Friedmann equation can be rewritten as
\begin{eqnarray}\label{defEz}
  E^2(z) &\equiv& \left(\frac{H}{H_0}\right)^2 \nonumber\\
   &=&
   \Omega_{{\rm rad}0}(1+z)^4+ \Omega_{{\rm mat}0}(1+z)^3+
       \Omega_{K0}(1+z)^{2}+\Omega_{\Lambda0}\ ,
\end{eqnarray}
with $\Omega_{K0}=1-\Omega_{\rm rad 0}- \Omega_{\rm mat 0}-
\Omega_{\Lambda0}$. All background observables, such as the
luminosity distance, the angular distance,\ldots, are functions of
$E(z)$ and are thus not independent.

Besides this background description, the $\Lambda$CDM also
accounts for an understanding of the large scale structure of our
universe (galaxy distribution, cosmic microwave background
anisotropy) by using the theory of cosmological perturbations at
linear order. In particular, in the sub-Hubble regime, the growth
rate of the density perturbation is also a function of $E(z)$.

One must, however, extend this minimal description by a primordial
phase in order to solve the standard cosmological problems
(flatness, horizon...). In our reference model, we assume that
this phase is described by an inflationary period during which the
expansion of the universe is almost-exponentially accelerated. In
such a case, the initial conditions for the gravitational dynamics
that will lead to the large scale structure are also determined so
that our model is completely predictive. We refer to the chapter~8
of Peter and Uzan (2005) for a detailed description of these
issues that are part of our cosmological model but not directly
related to our actual discussion.

In this framework, the dark energy is well defined and reduces to
a single number equivalent to a fluid with equation of state
$w=P/\rho=-1$. This model is compatible with all astronomical data
which roughly indicates that
$$
 \Omega_{\Lambda0} \simeq 0.73,\qquad
 \Omega_{\rm mat0} \simeq 0.27,\qquad
 \Omega_{\Lambda0} \simeq 0.
$$

\subsection{The cosmological constant problem}\label{sec116}

This model is theoretically well-defined, observationally
acceptable, phenomenologically simple and economical. From the
perspective of general relativity the value of $\Lambda$ is
completely free and there is no argument allowing us to fix it, or
equivalently, the length scale $\ell_\Lambda =
|\Lambda_0|^{-1/2}$, where $\Lambda_0$ is the astronomically
deduced value of the cosmological constant. Cosmology roughly
imposes that
$$
 |\Lambda_0|\leq H_0^2\,
 \Longleftrightarrow\,
 \ell_\Lambda \leq H_0^{-1} \sim 10^{26}\,\mathrm{m}
 \sim 10^{41}\,\mathrm{GeV}^{-1}\ .
$$
In itself this value is no problem, as long as we only consider
classical physics. Notice however that it is disproportionately
large compared  to the natural scale fixed by the Planck length
\begin{equation}
 \ell_\Lambda > 10^{60}\ell_{\rm P}\,\Longleftrightarrow\,
 \frac{\Lambda_0}{M_{\rm Pl}^2} < 10^{-120}\,\Longleftrightarrow\,
 \rho_{\Lambda_0}<10^{-120}M_{\rm Pl}^4\sim10^{-47}\,\mathrm{GeV}^4\
 ,
\end{equation}
when expressed in terms of energy density.

The main problem arises from the interpretation of the
cosmological constant. The local Lorentz invariance of the vacuum
implies that its energy-momen\-tum tensor must take the
form~(Zel'dovich, 1988) $\langle T^{\mathrm{vac}}_{\mu\nu}\rangle
= -\langle\rho\rangle g_{\mu\nu}$, that is equivalent to the one
of a cosmological constant. From the quantum point of view, the
vacuum energy receives a contribution of the order of
\begin{equation}
 \langle\rho\rangle^{\mathrm{EW}}_{\rm vac} \sim (200\,\mathrm{GeV})^4\ ,\qquad
 \langle\rho\rangle^{\mathrm{Pl}}_{\rm vac} \sim (10^{18}\,\mathrm{GeV})^4\ ,
\end{equation}
arising from the zero point energy, respectively fixing the cutoff
frequency of the theory to the electroweak scale or to the Planck
scale. This contribution implies a disagreement of respectively 60
to 120 orders of magnitude with astronomical observations!

This is the cosmological constant problem~(Weinberg, 1989). It
amounts to understanding why
\begin{equation}
 |\rho_{\Lambda_0}| = |\rho_\Lambda + \langle\rho\rangle_{\rm vac}|<10^{-47}\,\mathrm{GeV}^4
\end{equation}
or equivalently,
\begin{equation}
 |\Lambda_0| = |\Lambda + 8\pi G\langle\rho\rangle_{\rm vac}|<
 10^{-120}M_{\rm Pl}^2\ ,
\end{equation}
i.e. why $\rho_{\Lambda_0}$ is so small today, but non-zero.

Today, there is no known solution to this problem and two
approaches have been designed. One the one hand one sticks to this
model and extend the cosmological model in order to explain why we
observe a so small value of the cosmological constant~(Garriga and
Vilenkin, 2004 ; Carr and Ellis, 2008). We shall come back on this
approach later. On the other hand, one hopes that there should
exist a physical mechanism to exactly cancel the cosmological
constant and looks for another mechanism to explain the observed
acceleration of the Universe.

\subsection{The equation of state of dark energy}

The equation of state of the dark energy is obtained from the
expansion history, assuming the standard Friedmann equation. It is
thus given by the general expression~(Martin \etal, 2006)
\begin{equation}\label{refwde}
 3\Omega_\de w_\de =-1+\Omega_K + 2q,
\end{equation}
$q$ being the deceleration parameter,
\begin{equation}
 q\equiv-\frac{a\ddot a}{\dot a^2}=-1+\frac{1}{2}(1+z)\frac{\dd\ln H^2}{\dd z}.
\end{equation}
This expression~(\ref{refwde}) does not assume the validity of
general relativity or any theory of gravity but gives the relation
between the dynamics of the expansion history and the property of
the matter that would lead to this acceleration if general
relativity described gravity. Thus, the equation of state, as
defined in Eq.~(\ref{refwde}), reduces to the ratio of the
pressure, $P_\de$, to the energy density $\rho_\de$ of an
effective dark energy fluid under this assumption only, that is if
\begin{eqnarray}
 H^2 &=& \frac{8\pi G}{3}(\rho+\rho_\de) - \frac{K}{a^2},\\
 \frac{\ddot a}{a} &=& -\frac{4\pi G}{3}(\rho+\rho_\de+3P+3P_\de).
\end{eqnarray}
All the background information about dark energy is thus
encapsulated in the single function $w_\de(z)$. Most observational
constraints on the dark energy equation of states refer to this
definition.

\section{Modifying the minimal $\Lambda$CDM}\label{sec2}

The {\it Copernican principle} implies that the spacetime metric
reduces to a single function, the scale factor $a(t)$ that can be
Taylor expanded as $a(t)=a_0+H_0(t-t_0) - \frac{1}{2} q_0 H_0^2
(t-t_0)^2 +\ldots$. It follows that the conclusions that the
cosmic expansion is accelerating ($q_0<0$) does not involve any
hypothesis about the theory of gravity (other than the one that
the spacetime geometry can be described by a metric) or the matter
content, as long as this principle holds.

The assumption that the Copernican principle holds, and the fact
that it is so central in drawing our conclusion on the
acceleration of the expansion, splits our investigation into two
avenues. Either we assume that the Copernican principle holds and
we have to modify the laws of fundamental physics or we abandon
the Copernican principle, hoping to explain dark energy without
any new physics but at the expense of living in a particular place
in the Universe. While the first solution is more orthodox from a
cosmological point of view, the second is indeed more conservative
from a physical point of view. It will be addressed in
\S~\ref{sec125}. We are thus in front of a choice between
``simple'' cosmological solutions with new physics and more
involved cosmological solutions of standard physics.

This section focuses on the first approach. If general relativity
holds then Eq.~(\ref{fried2}) tells us that the dynamics has to be
dominated by a dark energy fluid with $w_\de<-\frac13$ for the
expansion to be accelerated. The simplest solution is indeed the
cosmological constant $\Lambda$ for which $w_\de=-1$ and which is
the only model not introducing new degrees of freedom.

\subsection{General classification of physical models}

\subsubsection{General Relativity}

Einstein's theory of gravity relies on two independent hypotheses.

First, the theory rests on the Einstein equivalence principle,
which includes the universality of free fall, the local position
and local Lorentz invariances in its weak form (as other metric
theories) and is conjectured to satisfy it in its strong form. We
refer to Will (1981) for a detailed explanation of these
principles and their implications. The weak equivalence principle
can be mathematically implemented by assuming that all matter
fields are minimally coupled to a single metric tensor
$g_{\mu\nu}$. This metric defines the length and times measured by
laboratory clocks and rods so that it can be called the {\it
physical metric}. This implies that the action for any matter
field, $\psi$ say, can be written as $S_{\rm
matter}[\psi,g_{\mu\nu}]$. This so-called {\it metric coupling}
ensures in particular the validity of the universality of
free-fall.

The action for the gravitational sector is given by the
Einstein-Hilbert action
\begin{equation}
 S_{\rm gravity} = \frac{c^3}{16\pi G}\int \dd^4x\sqrt{-g_*}R_*,
\end{equation}
where $g^*_{\mu\nu}$ is a massless spin-2 field called the {\it
Einstein metric}. The second hypothesis states that both metrics
coincide
$$
 g_{\mu\nu} = g^*_{\mu\nu}.
$$

The underlying physics of our reference cosmological model (i.e.
hypotheses H1 and H2) is thus described by the action
\begin{equation}\label{action_reference}
 S_{\rm gravity} = \frac{c^3}{16\pi G}\int \dd^4x\sqrt{-g}(R-2{\bf\Lambda})
 + \!\!\!\!\! \!\!\sum_{{\rm standard\, model}+{\rm\bf CDM}}\!\!\!S_{\rm
 matter}[\psi_i,g_{\mu\nu}],
\end{equation}
which includes all known matter fields plus two unknown components
(in bold face).

\subsubsection{Local experimental constraints}\label{sec1222}

The assumption of a metric coupling is well tested in the Solar
system. First it implies that all non-gravitational constants are
spacetime independent, which have been tested to a very high
accuracy in many physical systems and for various fundamental
constants~(Uzan, 2003; Uzan, 2004; Uzan and Leclercq, 2008), e.g.
at the $10^{-7}$ level for the fine structure constant on time
scales ranging to 2-4 Gyrs. Second, the isotropy has been tested
from the constraint on the possible quadrupolar shift of nuclear
energy levels~(Prestage \etal, 1985; Chupp \etal, 1989; Lamoreaux
\etal, 1986) proving that matter couples to a unique metric tensor
at the $10^{-27}$ level. Third, the universality of free fall of
test bodies in an external gravitational field at the $10^{-13}$
level in the laboratory~(Baessler \etal, 1999; Adelberger, \etal
2001). The Lunar Laser ranging experiment~(Williams \etal, 2004),
which compares the relative acceleration of the Earth and Moon in
the gravitational field of the Sun, also probe the strong
equivalence principle at the $10^{-4}$ level. Fourth, the Einstein
effect (or gravitational redshift) states that two identical
clocks located at two different positions in a static Newton
potential $U$ and compared by means of electromagnetic signals
shall exhibit a difference in clock rates of $1+[U_1-U_2]/c^2$,
where $U$ is the gravitational potential. This effect has been
measured at the $2\times10^{-4}$ level~(Vessot and Levine, 1978).

The parameterized post-Newtonian formalism (PPN) is a general
formalism that introduces 10 phenomenological parameters to
describe any possible deviation from general relativity at the
first post-Newtonian order~(Will, 1981). The formalism assumes
that gravity is described by a metric and that it does not involve
any characteristic scale. In its simplest form, it reduces to the
two Eddington parameters entering the metric of the Schwartzschild
metric in isotropic coordinates
$$
 g_{00} = - 1 + \frac{2Gm}{rc^2} -
 2\beta^\ppn\left(\frac{2Gm}{rc^2}\right)^2,
 \qquad
 g_{ij} = \left(1+2\gamma^\ppn\frac{2Gm}{rc^2}\right)\delta_{ij}.
$$
Indeed, general relativity predicts $\beta^\ppn=\gamma^\ppn=1$.
These two phenomenological parameters are constrained (1) by the
shift of the Mercury perihelion~(Shapiro \etal, 1990) which
implies that $|2\gamma^\ppn-\beta^\ppn-1|<3\times10^{-3}$, (2) the
Lunar laser ranging experiments~(Williams \etal, 2004) which
implies $|4\beta^\ppn-\gamma^\ppn-3|=(4.4\pm4.5)\times10^{-4}$ and
(3) by the deflection of electromagnetic signals which are all
controlled by $\gamma^\ppn$. For instance the very long baseline
interferometry~(Shapiro \etal, 2004) implies that
$|\gamma^\ppn-1|=4\times10^{-4}$ while the measurement of the time
delay variation to the Cassini spacecraft~(Bertotti \etal, 2003)
sets $\gamma^\ppn-1=(2.1\pm2.3)\times10^{-5}$.

The PPN formalism does not allow to test finite range effects that
could be caused e.g. by a massive degree of freedom. In that case
one expects a Yukawa-type deviation from the Newton potential,
$$
 V=\frac{Gm}{r}\left(1+\alpha\hbox{e}^{-r/\lambda}\right),
$$
that can be probed by ``fifth force'' experimental searches.
$\lambda$ characterizes the range of the Yukawa deviation while
its strength $\alpha$ may also include a
composition-dependence~(Uzan, 2003). The constraints on
$(\lambda,\alpha)$ are summarized in Hoyle \etal (2004) which
typically shows that $\alpha<10^{-2}$ on scales ranging from the
millimeter to the Solar system size.

In general relativity, the graviton is massless. One can however
give it a mass, but this is very constrained. In particular,
around a Minkowski background, the mass term must have the very
specific form of the Pauli-Fierz type in order to avoid ghosts
(see below for a more precise definition) to be excited. This mass
term is however inconsistent with Solar system constraints because
there exists a discontinuity~(van Dam and Veltman, 1970; Zakharov,
1970) between the case of a strictly massless graviton and a very
light one. In particular, such a term can be ruled out from the
Mercury perihelion shift.

General relativity is also tested  with pulsars~(Damour, and
Esposito-Far\`ese, 1998; Esposito-Far\`ese, 2005) and in the
strong field regime~(Psaltis, 2008). For more details we refer to
Will (1981), Damour and Lilley (2008) and Turyshev (2008).
Needless to say that any extension of general relativity has to
pass these constraints. However, deviations from general
relativity can be larger in the past, as we shall see, which makes
cosmology an interesting physical system to extend these
constraints.

\subsubsection{Universality classes}

There are many possibilities to extend this minimal physical
framework. Let us start by defining universality classes~(Uzan,
2007) by restricting our discussion to field theories. This has
the advantage to identify the new degrees of freedom and their
couplings.

  \begin{figure}
    \centering
    \includegraphics[width=12cm]{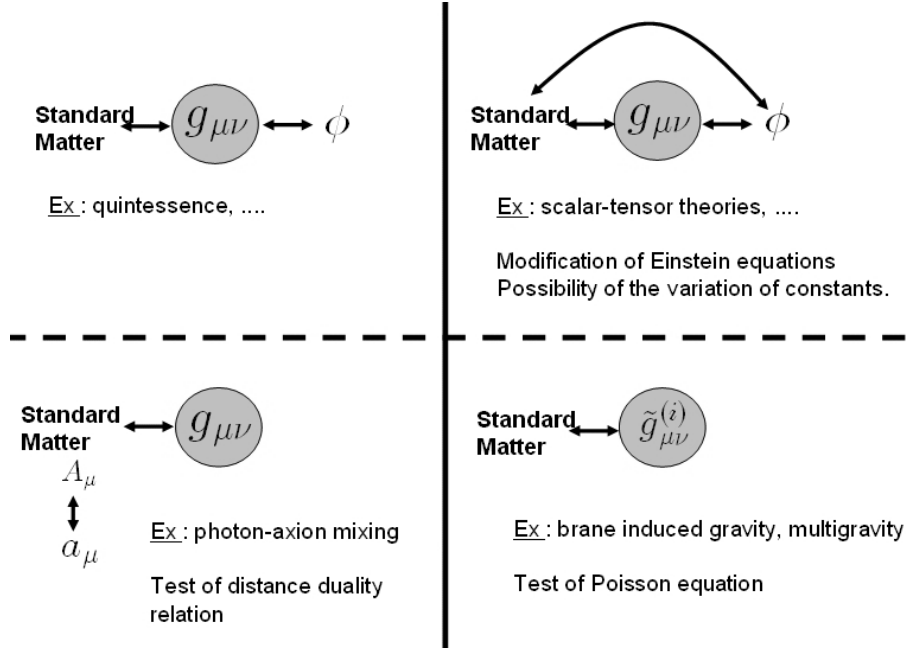}
    \caption{Summary of the different classes of physical dark energy models. As discussed in the text,
 various tests can be designed to distinguish between them. The classes differ according
 to the nature of the new degrees of freedom and their couplings.
 Left column accounts for models
 where gravitation is described by general relativity while right column models
 describe a modification of general relativity. In the upper line classes, the new fields
 dominate the matter content of the universe at low redshift.
 Upper-left models (class A) consist of
 models in which a new kind of gravitating matter is introduced. In the upper-right
 models (class C), a light field induces a long-range force so that gravity is
 not described by a massless spin-2 graviton only. In this class, Einstein equations
 are modified and there may be a variation of the fundamental
 constants. The lower-right models (class D) correspond to models in which there
 may exist an infinite number of new degrees of freedom, such as in some class of braneworld
 scenarios. These models predict a modification of the Poisson
 equation on large scales. In the last class (lower-left, class B), the
 distance duality relation may be violated. From Uzan (2007).}
    \label{fig2}
  \end{figure}

The first two classes assume that gravitation is well described by
general relativity and introduce new degrees of freedom beyond
those of the standard model of particle physics. This means that
one adds a new term $S_{\rm de}[\psi;g_{\mu\nu}]$ in the
action~(\ref{action_reference}) while keeping the Einstein-Hilbert
action and the coupling of all the fields (standard matter and
dark matter) unchanged. They are:
\begin{enumerate}
 \item[1.]\underline{Class A} consists of models in which the acceleration is
 driven by the gravitational effect of the new fields.
 They thus must have an equation of state smaller than $-\frac13$.
 They are not coupled to the standard matter fields
 or to dark matter so that one is adding a new sector
 $$
 S_{\rm de}[\phi;g_{\mu\nu}]
 $$
 to the action~(\ref{action_reference}), where $\phi$ stands for the
 dark energy field (not necessarily a scalar field).
 Standard examples include {\it quintessence}
 models~(Wetterich, 1988; Ratra and Peebles, 1988)
 which invoke a canonical scalar
 field slow-rolling today, {\it solid dark matter}
 models~(Battye \etal, 1999) induced by frustrated topological defects networks,
 {\it tachyon} models~(Sen, 1999), {\it Chaplygin gas}~(Kamenshchik \etal, 2001) and
 {\it K-es\-sence}~(Armendariz-Picon \etal, 2000;
 Chiba \etal, 2000)
 models invoking scalar fields with a non-canonical kinetic term.
 \item[2.]\underline{Class B} introduces new fields which do not dominate the matter
 content so that they do not change the expansion rate of the
 universe. They are thus not required to have an
 equation of state smaller than $-\frac13$. These
 fields are however coupled to photons and thus affect the observations.
 An example~(Csaki \etal, 2002; Deffayet \etal, 2002) is provided
 by {\it photon-axion oscillations}
 which aims at explaining the dimming of supernovae not by an accelerated
 expansion but by the
 fact that part of the photons has oscillated into invisible
 axions. In that particular case, the
 electromagnetic sector is modified according to
 $$
 S_{\rm em}[A_\mu;g_{\mu\nu}] \rightarrow S_{\rm em}[A_\mu, a_\mu;g_{\mu\nu}].
 $$
 A specific signature of these models would be a violation of the distance duality
 relation (see \S~\ref{sec1331}).
\end{enumerate}

Then come models with a modification of general relativity. Once
such a possibility is considered, many new models arise~(Will,
1981). They are:
\begin{enumerate}
\item[3.]\underline{Class C} includes models in which a finite
number of new fields are
 introduced. These fields couple to the standard model fields and some of them dominate the matter
 content (at least at late time). This is the case in particular of scalar-tensor theories in
which a
 scalar field couples universally and leads to the class of extended quintessence
 models, chameleon models or $f(R)$ models depending on
 the choice of the coupling
 function and potential (see \S ~\ref{sec124}). For these models, one has a new sector
 $$
 S_\varphi[\varphi;g_{\mu\nu}]
 $$
 and the couplings of the matter fields will be modified according to
 $$
 S_{\rm matter}[\psi_i;g_{\mu\nu}]\rightarrow S_{\rm matter}
 [\psi_i;A^2_i(\varphi)g_{\mu\nu}].
 $$
 If the coupling is not universal, a signature may be the
 variation of fundamental constants and a violation of the universality of free fall.
 This class also
 offers the possibility to enjoy $w_\de<-1$ with a well-defined field
 theory and includes models in which a scalar
 field couples differently to standard matter field and dark matter.
\item[4.]\underline{Class D} includes more drastic modifications
of general relativity with e.g. the
 possibility to have more types of gravitons (massive or not and most probably an
 infinite number of them). This is for instance the case of models involving extra-dimensions such as
 e.g. multi-brane models~(Gregory \etal, 2000), multigravity~(Kogan \etal, 2000), brane
 induced gravity~(Dvali \etal, 2000) or simulated
 gravity~(Carter \etal, 2001). In these cases, the new fields modified the gravitational
 interaction on large scale but do not necessarily dominate the matter content
 of the universe. Some of these models may also offer the possibility to mimic an equation of
 state $w_\de<-1$.
\end{enumerate}

These various modifications, summarized on Fig.~\ref{fig2} can
indeed be combined to get more exotic models.

\subsubsection{``Modified gravity'' vs new matter}

The different models in the literature are often categorized as
``modified gravity'' or ``new matter''. This distinction may
however be subtle.

First, we shall define {\it gravity} as the long range force that
cannot be screened. We are used to describe this interaction by
general relativity so that it is associated with a massless spin-2
graviton. In our view, gravity cannot be modified but only its
description, i.e. general relativity. As an example, scalar-tensor
theories (see \S~\ref{sec124}) extend general relativity by a
spin-0 interaction which can be long-range according to the mass
of the scalar field. In this case, the interaction is even
universal so that it does not imply any violation of the weak
Einstein equivalence principle.

Note also that whatever the model, it requires the introduction of
new fields beyond those of the standard model. The crucial
difference is that in models with ``new matter'' (e.g. class A),
the amount of dark energy is imposed by initial conditions and its
gravitational effect induces the acceleration of the Universe. In
a ``modified gravity'' model (e.g. classes C and D) the standard
matter and cold dark matter generate an effective dark energy
component. The acceleration may thus be a consequence of the fact
that the gravitational interaction is weaker than expected on
large scales. But, it may be that the energy density of the new
field also dominates the dynamics but still be determined by the
energy density of the standard field.

\subsection{Modifying General Relativity}

\subsubsection{In which regime?}\label{reg}

Before investigating gravity beyond general relativity, let us try
to sketch the regimes in which these modifications may (or shall)
appear. We can distinguish the following regimes.
\begin{itemize}
 \item{\it Weak-strong field regimes} can be characterized by the
 amplitude of the gravitational potential. For a spherical static
 spacetime, $\Phi=GM/rc^2$. It is of order of $\Phi_\odot\sim 2\times10^{-6}$ at
 the surface of the Sun and equal to $\frac12$ for a black-hole.
 \item{\it Small-large distances}. Such modifications can be induced
 by a massive degree of freedom that will induce a Yukawa like
 coupling. While constrained on the size of the solar system, we
 have no constraints on scales larger that $10 h^{-1}$~Mpc.
 \item{\it Low-high acceleration regimes} are of importance to
 discuss galaxy rotation curves and (galactic) dark matter, as suggested by
 the MOND phenomenology~(Milgrom, 1983). In particular the kind of
 modification of the gravitation theory that could account
 for the dark matter cannot occur at a characteristic distance because of
 the Tully-Fischer law.
 \item{\it Low-high curvature regimes} will distinguish the
 possible extensions of the Einstein-Hilbert action. For instance
 a quadratic term of the form $\alpha R^2$ becomes significant
 compared to $R$ when $GM/r^3c^2\gg\alpha^{-1}$ even if $\Phi$
 remains small. In the Solar system
 $R_\odot\sim4\times10^{-28}\hbox{cm}^{-2}$.
\end{itemize}

In cosmology, we can suspect various possible regimes in which to
modify general relativity. The dark matter problem can be
accounted for by a modification of Newton gravity below the
typical acceleration $a_0\sim10^{-8}\hbox{cm.s}^{-2}$. It follows
that the regime for which a dark matter component is required can
be characterized by
\begin{equation}\label{emond}
 \Phi R < a_0^2\sim3\times10^{-31}R_\odot.
\end{equation}
Concerning the homogeneous universe, one can sort out from the
Friedmann equations that
\begin{equation}\label{eFL}
 R_{\rm FL}(z) = 3H_0^2[\Omega_{\mat0}(1+z)^3+4\Omega_{\Lambda0}],
\end{equation}
from which we deduce that $R_{\rm FL}\sim 10^{-5}R_\odot$ at the
time of nucleosynthesis, $R_{\rm FL}\sim 10^{-20}R_\odot$ at the
time of decoupling and $R_{\rm FL}\sim 10^{-28}R_\odot$ at $z=1$.
The curvature scale associated to a cosmological constant is
$R_\Lambda=\frac16\Lambda$ and the cosmological constant (or dark
energy) problem corresponds to a low curvature regime,
\begin{equation}\label{ede}
 R<R_\Lambda\sim1.2\times10^{-30}R_\odot.
\end{equation}
The fact that the limits~(\ref{emond}) and~(\ref{ede}) intersect
illustrates the coincidence problem, that is $a_o\sim cH_0$ and
$\Omega_{\mat0}\sim\Omega_{\Lambda0}$. Note that both arise on
curvature scales much smaller than those probed in the solar
system.

Let us now turn to the cosmological perturbations. The
gravitational potential at the time of the decoupling
($z\sim10^3$) is of the order of $\Phi\sim10^{-5}$. During the
matter era, the Poisson equation imposes that
$\Delta\Phi\propto\delta\rho_\mat a^2$ which is almost constant.
It follows that we never expect a potential larger than
$\Phi\sim10^{-5}$ on cosmological scales. We are thus always in a
weak field regime. The characteristic distance scale is fixed by
the Hubble radius $c/H_0$. The curvature perturbation associated
with the large scale structures is, in the linear theory, of the
order
$$
 \delta R =\frac{6}{a^2}\Delta\Phi\sim 3H_0^2\Omega_{\mat0}(1+z)^3
 \delta_\mat(z).
$$
Since at redshift zero, $\langle\delta_\mat^2\rangle = \sigma_8
\sim 1$ in a ball of radius of 8 Mpc, we conclude that
$\langle\delta R^2\rangle^{1/2}\sim 3H_0^2\Omega_{\mat0}\sigma_8$
while $R_{\rm FL}=3H_0^2\Omega_{\mat0}$ if $\Lambda=0$. This means
that the curvature perturbation becomes of the order of the
background curvature at a redshift $z\sim0$, even if we are still
in the weak field limit. This implies that the effect of the large
scale structures on the background dynamics may be non-negligible.
This effect has been argued to be at the origin of the
acceleration of the universe~(Ellis and Buchert, 2005; Ellis,
2008) but no convincing formalism to describe this backreaction
has been constructed yet. Note that in this picture, the onset of
the acceleration phase will be determined by the amplitude of the
initial power spectrum.

In conclusion, to address the dark energy or dark matter problem
by a modification of general relativity, we are interested in
modifications on large scales (typically Hubble scales), low
acceleration (below $a_0$) or small curvature (typically
$R_\Lambda$).

\subsubsection{General constraints}

In modifying general relativity, we shall demand that the new
theory
\begin{itemize}
 \item {\it does not contain ghosts}, i.e. degrees of freedom with
 negative kinetic energy. The problem with such a ghost is that
 the theory would be unstable. In particular, the vacuum can
 decay in an arbitrary amount of positive energy (standard)
 gravitons
 whose energy would be balanced by negative energy ghosts.
 \item {\it has a Hamiltonian bounded from below}. Otherwise, the
 theory would be unstable, even if one cannot explicitly identify
 a ghost degree of freedom.
 \item the new degrees of freedom are not {\it tachyon}, i.e. do not
 have a negative mass.
 \item is {\it compatible with local tests} of deviation from general
 relativity, in particular in the Solar system described in
 \S~\ref{sec1222}.
\end{itemize}

Then, starting from the action~(\ref{action_reference}), we see
that we can either modify the Einstein-Hilbert action while
letting the coupling of all matter fields to the metric unchanged
or modify the coupling(s) in the matter action. The possibilities
are numerous~(Will, 1981; Esposito-Far\`ese and Bruneton, 2007;
Uzan, 2007) and we cannot start an extensive review of the models
here. We shall thus consider some examples that will illustrate
the constraints cited above, but with no goal of exhaustivity.

\subsubsection{Modifying the Einstein-Hilbert
action}\label{sechog}

Let us start with the example of higher order gravity models based
on the quadratic action (here we follow the very clear analysis of
Esposito-Far\`ese and Bruneton (2007) for our discussion)
\begin{equation}\label{EHquad}
 S_{\rm gravity} = \frac{c^3}{16\pi G}\int \dd^4x\sqrt{-g}
 \left[R+\alpha C^2_{\mu\nu\rho\sigma} + \beta R^2 + \gamma \hbox{GB} \right],
\end{equation}
where $C_{\mu\nu\rho\sigma}$ is the Weyl tensor and
$\hbox{GB}\equiv R^2_{\mu\nu\rho\sigma} -4R^2_{\mu\nu}+R^2$ is the
Gauss-Bonnet term. $\alpha$, $\beta$ and $\gamma$ are three
constants with dimension of an inverse mass square. Since GB does
not contribute to the local field equations of motion, we will not
consider it further. The action~(\ref{EHquad}) gives a
renormalisable theory of quantum gravity at all order provided
$\alpha$ and $\beta$ are non-vanishing~(Stelle, 1978). However,
such theories contain ghosts. This can be seen from the graviton
propagator which takes the form $1/(p^2+\alpha p^4)$. It can
indeed be decomposed in irreducible fractions as
$$
 \frac{1}{p^2+\alpha p^4}=
 \frac{1}{p^2}-\frac{1}{p^2+\frac{1}{\alpha}}.
$$
The first term is nothing but the standard propagator of the usual
massless graviton. The second term correspond to an extra-massive
degree of freedom with mass $\alpha^{-1}$ and its negative sign
indicates that it carries negative energy: it is a ghost. Moreover
if $\alpha$ is negative, this ghost is also a tachyon! The only
viable such modification arises from $\beta R^2$, which introduces
a massive spin-0 degree of freedom.

These considerations can be extended to more general theories
involving an arbitrary function of the metric invariants,
$f(R,R_{\mu\nu},$ $R_{\mu\nu\rho\sigma})$, which also
generically~(Hindawi \etal, 1996; Tomboulis, 1996) contain a
massive spin-2 ghost. They are thus not stable theories with the
exception of $f(R)$ theories, discussed in~\S~\ref{secfR}.

A possibility may seem to consider models designed such that their
second-order expansion never shows any negative energy kinetic
term. As recalled in Esposito-Far\`ese and Bruneton (2007) and
Woodard (2006), these models still exhibit instabilities the
origin of which can be related to a theorem by Ostrogradsky (1850)
showing that their Hamiltonian is generically not bounded from
below.

We summarized this theorem following the presentation by Woodard
(2006). Consider a Lagrangian depending on a variable $q$ and its
first two time derivatives $\mathcal{L}(q,\dot q,\ddot q)$ and
assume that it is not degenerate, i.e. that $\ddot q$ cannot be
eliminated by an integration by part. Then the definition
$p_2\equiv\partial\mathcal{L}/\partial\ddot q$ can be inverted to
get $\ddot q$ as a function $q$, $\dot q$ and $p_2$, $\ddot
q[q,\dot q, p_2]$, and the initial data must be specified by two
pairs of conjugate momenta defined by
$(q_1,p_1)\equiv(q,\partial\mathcal{L}/\partial\dot q
-\dd(\partial\mathcal{L}/\partial\ddot q)/\dd t)$ and
$(q_2,p_2)\equiv(\dot q,\partial\mathcal{L}/\partial\ddot q)$. The
Hamiltonian defined as $\mathcal{H}=p_1\dot q_1+p_2\dot
q_2-\mathcal{L}$ can be shown to be the generator of time
translations and the Hamilton equations which derive from
$\mathcal{H}$ are indeed equivalent to the Euler-Lagrange
equations derived from $\mathcal{L}$. In terms of $q_i$ and $p_i$,
the Hamiltonian takes the form
$$
 \mathcal{H} = p_1q_2 + p_2\ddot q[q_1,q_2,p_2] -
 \mathcal{L}(q_1,q_2,\ddot q[q_1,q_2,p_2]).
$$
This expression is however linear in $p_1$ so that the Hamiltonian
is not bounded from below and the theory is necessarily unstable.
Let us note that this constraint can be avoided by non-local
theories, that is if the Lagrangian depends on an infinite number
of derivatives, as e.g. string theory, even though its expansion
may look pathological.

\subsubsection{Modifying the matter action}\label{sec1234}

Many other possibilities, known as bi-metric theories of gravity,
arise if one assumes that $g_{\mu\nu}\not= g^*_{\mu\nu}$. Instead
one can postulate that the physical metric is a combination of
various fields, e.g.
$$
 g_{\mu\nu}[g_{\mu\nu}^*,\varphi,A_\mu,B_{\mu\nu},\ldots]
 =A^2(\varphi)\left[g_{\mu\nu}^*+ \alpha_1A_\mu A_\nu +
 \alpha_2g_{\mu\nu}^*g_*^{\alpha\beta}A_\alpha A_\beta+\ldots
 \right].
$$
As long as these new fields enter quadratically, their field
equation is generically of the form $\Box A= A T$ where $T$ is the
matter source. It follows that matter cannot generate them if
their background value vanishes. On the other hand, if their
background value does not vanish then these fields define a
preferred frame and Lorentz invariance is violated.

Such modifications have however drawn some attention, in
particular in the attempts of constructing a field theory
reproducing the MOND phenomenology~(Milgrom, 1983). In particular,
in order to increase light deflection in scalar-tensor theories of
gravity, a {\it disformal coupling}~(Bekenstein, 1993), $
g_{\mu\nu} = A^2(\varphi)g_{\mu\nu}^*+ B(\varphi)
\partial_\mu\varphi \partial_\nu\varphi$, was introduced. It was
generalized to {\it stratified theory}~(Sanders, 1997). by
replacing the gradient of the scalar field by a dynamical unit
vector field ($g_{\mu\nu}^*A^\mu A^\nu=-1$), $g_{\mu\nu}
=A^2(\varphi)g_{\mu\nu}^* + B(\varphi)A_\mu A_\nu$. This is at the
basis of the TeVeS theory proposed by Bekenstein~(Bekenstein,
2004). The mathematical consistency and the stability of these
field theories were investigated in depth in the excellent
analysis of Esposito-Far\`ese and Bruneton (2007). It was shown
that no present theory passes all available experimental
constraints while being stable and admitting a well-posed Cauchy
problem.

Esposito-Far\`ese and Bruneton (2007) also notice that while
couplings of the form $g_{\mu\nu}[g_{\mu\nu}^*,R_{\mu\nu}^*,
R_{\mu\nu\alpha\beta}^{*},\ldots]$ seem to lead to well-defined
theories in vacuum (in particular) when linearizing around a
Minkowsky background, they are unstable inside matter, because the
Ostrogradsky theorem strikes back.

The case in which only a scalar partner,
$g_{\mu\nu}=A^2(\varphi)g_{\mu\nu}^*$, is introduced leads to
consistent field theories and is the safest way to modify the
matter coupling. We shall discuss these scalar-tensor theories of
gravity in \S~\ref{sec124}.

\subsubsection{Higher-dimensional theories}

Higher-dimensional models of gravity, among which string
theory~(see e.g. Damour and Lilley (2008)) predict non-metric
coupling as those discussed in the previous section. Many scalar
fields, known as {\it moduli}, appear in the dimensional reduction
to four dimensions.

As a simple example, let us consider a five-dimensional spacetime
and assume that gravity is described by the Einstein-Hilbert
action
\begin{equation}\label{13.R5}
 S = \frac{1}{12\pi^2 G_5} \int \bar{R}\sqrt{|\bar{g}|}\,\dd^5 x,
\end{equation}
where we denote by a bar quantities in 5 dimensions to distinguish
them with the analogous quantities with no bar in 4 dimensions.
The aim is to determine the  independent elements of the metric
$g_{AB}$, which are 15 in five dimensions. We decompose the metric
into a symmetric tensor part $g_{\mu\nu}$, with $10$ independent
components, a vector part, $A_\alpha$, with four components and
finally a scalar field, $\phi$, to complete the counting of the
number of degrees of freedom ($15= 10 + 4 +1$). The metric is thus
decomposed as
\begin{equation}\label{13.gAB}
  \bar{g}_{AB} =\left(
  \begin{array}{cc}
  g_{\mu\nu} + \frac{1}{M^2} \phi^2 A_\mu A_\nu & \frac{1}{M}\phi^2
  A_\mu
  \\
  \frac{1}{M} \phi^2 A_\nu & \phi^2
  \end{array}\right)
  ,
\end{equation}
where the different components depend {\sl a priori} both on the
usual space-time coordinates $x^\alpha$ and the coordinate in the
extra-dimension $y$.  The constant $M$ has dimensions of a mass,
so that $A_\alpha$ also has dimensions of mass, whereas the scalar
field $\phi$ is here dimensionless. Finally, while capital latin
indices vary in the entire 5-dimensional space-time, $A,B =
0,\cdots,4$, greek indices span the 4-dimensional space-time,
namely $\mu,\nu = 0,\cdots,3$. Compactifying on a circle and
assuming that none of the variables depends on the transverse
direction $y$ ({\it cylinder condition}), the action~(\ref{13.R5})
reduces to the four-dimensional action
\begin{equation}\label{13.R4}
 S = \frac{1}{16\pi G} \int \dd^4x\sqrt{-g} \phi \left( R
 -\frac{\phi^2}{4M^2} F_{\alpha\beta} F^{\alpha\beta}\right),
\end{equation}
where $F_{\alpha\beta}\equiv \partial_\alpha A_\beta -
\partial_\beta A_\alpha$ and where we have set
$$
 G = \frac{3\pi\bar{G}_5}{4 V_{(5)}},
$$
and factored out the finite volume of the fifth dimension,
$V_{(5)}=\int \dd y$. The scalar field couples explicitly to the
kinetic term of the vector field. It can be checked that this
coupling cannot be eliminated by a redefinition of the metric,
whatever the function $A(\phi)$: this is the well-known conformal
invariance of electromagnetism in four dimensions. Such a term
induces a variation of the fine structure constant as well as a
violation of the universality of free-fall~(Uzan, 2003). Such
dependencies of the masses and couplings are generic for
higher-dimensional theories and in particular string theory.

The cylinder condition is justified as long as we consider the
fifth dimension to be topologically compact with the topology of a
circle. In this case, all the fields which are defined in this
space, i.e. the four-dimensional metric $g_{\mu\nu}$, the vector
$A_\alpha$ and the dilaton $\phi$, and any additional matter
fields that the theory should describe, are periodic functions of
the extra-dimension and can therefore be expanded into Fourier
modes. The radius $R$ of this dimension then turns out to be
naturally $R\sim M^{-1}$. For large enough $M$, the radius is too
small to have observable consequences: to be sensitive to the
fifth dimension, the energies involved must be comparable to $M$.
Decomposing all the fields in Fourier modes a e.g.
\begin{equation}\label{13.Fourier}
\phi\left(x_\mu,y\right) = \sum_{n=-\infty}^{+\infty} \phi_n
  \left(x_\mu\right) \hbox{e}^{i n M y}, \qquad \hbox{with} \qquad
  \phi_{-n} = \phi^\star_n
\end{equation}
($\phi$ real), we conclude that the four-dimensional theory will
also contain a infinite tower of modes of increasing mass.

While these tree-level predictions of string theory are in
contradiction with experimental constraints, many mechanisms can
reconcile it with experiment. In particular, it has been claimed
that quantum loop corrections to the tree-level action may modify
the coupling in such a way that it has a minimum~(Damour and
Polyakov, 1994). The scalar field can thus be attracted toward
this minimum during the cosmological evolution so that the theory
is attracted toward general relativity. Another possibility is to
invoke an environmental dependence, as can be implemented in
scalar-tensor theories by the chameleon mechanism~(Khoury and
Weltman, 2004) which invokes a potential with a minimum not
coinciding with the one of the coupling function.

In higher dimensions, the Einstein-Hilbert action can also be
modified by adding the Gauss-Bonnet term GB since it does not
enter the field equations only in four dimensions. The
$D$-dimensional Einstein-Hilbert action can then be modified to
include a term of the form $\alpha$GB. In particular, it is the
case in the low-energy limit of heterotic string theory~(Gross and
Sloan, 1987). In various configurations, in particular with
branes, it has been argued that the Gauss-Bonnet invariant can
also couple to a scalar field~(Amendola \etal, 2006), i.e.
$\alpha(\varphi)$GB. As long as the modification is linear in GB,
it is ghost-free.

In the context of braneworld, it was shown that some models with
infinite volume extra-dimension can produce a modification of
general relativity leading to an acceleration of the expansion. In
the DGP model~(Dvali \etal, 2000), one considers beside the
5-dimensional Einstein-Hilbert a 4-dimensional term induced on the
brane
\begin{equation}\label{13.dgp}
 S = \frac{M_5^2}{2} \int \bar{R_5}\sqrt{|\bar{g_5}|}\,\dd^5x +
 \frac{M_4^2}{2} \int R_4\sqrt{|g_4|}\,\dd^4x.
\end{equation}
There is a competition between these two terms and the
five-dimensional term dominates on scales larger than
$r_c=M_4^2/2M_5^3$.  The existence or absence of ghost in this
class of models is still under debate. Some of these
models~(Deffayet, 2005) have also been claimed to describe massive
gravitons without being plagued by the van Dam-Veltman-Zakharov
discontinuity (see \S~\ref{sec1222}).

As a conclusion, higher-dimensional models offer a rich variety of
possibilities among which some may be relevant to describe a
modification of general relativity on large scales.

\subsection{Example: scalar-tensor theories}\label{sec124}

As discussed in \S~\ref{sec1234}, the case in which only a scalar
partner to the graviton is introduced leads to consistent field
theories and is the safest way to modify the matter coupling.

\subsubsection{Formulation}

In  scalar-tensor theories, gravity is mediated not only by a
massless spin-2 graviton but also by a spin-0 scalar field that
couples universally to matter fields (this ensures the
universality of free fall). In the Jordan frame, the action of the
theory takes the form
\begin{eqnarray}\label{actionJF}
  S &=&\int \frac{\dd^4 x }{16\pi G_*}\sqrt{-g}
     \left[F(\varphi)R-g^{\mu\nu}Z(\varphi)\varphi_{,\mu}\varphi_{,\nu}
        - 2U(\varphi)\right]\nonumber\\
        &&\qquad\qquad+ S_{\rm matter}[\psi;g_{\mu\nu}]
\end{eqnarray}
where $G_*$ is the bare gravitational constant. This action
involves three arbitrary functions ($F$, $Z$ and $U$) but only two
are physical since there is still the possibility to redefine the
scalar field. $F$ needs to be positive to ensure that the graviton
carries positive energy. $S_{\rm matter}$ is the action of the
matter fields that are coupled minimally to the metric
$g_{\mu\nu}$. In the Jordan frame, the matter is universally
coupled to the metric so that the length and time as measured by
laboratory apparatus are defined in this frame.

It is useful to define an Einstein frame action through a
conformal transformation of the metric
\begin{equation}\label{jf_to_ef}
 g_{\mu\nu}^* = F(\varphi)g_{\mu\nu}.
\end{equation}
In the following all quantities labelled by a star (*) will refer
to Einstein frame. Defining the field $\varphi_*$ and the two
functions $A(\varphi_*)$ and $V(\varphi_*)$ (see e.g.
Esposito-Far\`ese and Polarski, 2001) by
\begin{eqnarray}
 \left(\frac{\dd\varphi_*}{\dd\varphi}\right)^2
              &=& \frac{3}{4}\left(\frac{\dd\ln F(\varphi)}{\dd\varphi}\right)^2
                  +\frac{1}{2F(\varphi)}\label{jf_to_ef1}\\
 A(\varphi_*) &=& F^{-1/2}(\varphi)\label{jf_to_ef2}\\
 2V(\varphi_*)&=& U(\varphi) F^{-2}(\varphi)\label{jf_to_ef3},
\end{eqnarray}
the action (\ref{actionJF}) reads as
\begin{eqnarray}
 S &=& \frac{1}{16\pi G_*}\int \dd^4x\sqrt{-g_*}\left[ R_*
        -2g_*^{\mu\nu} \partial_\mu\varphi_*\partial_\nu\varphi_*
        - 4V(\varphi_*)\right]\nonumber\\
   && \qquad + S_{\rm matter}[A^2(\varphi_*)g^*_{\mu\nu};\psi].
\end{eqnarray}
The kinetic terms have been diagonalised so that the spin-2 and
spin-0 degrees of freedom of the theory are perturbations of
$g^*_{\mu\nu}$ and $\varphi_*$ respectively.

In this frame, the field equations take the form
\begin{eqnarray}
 G^*_{\mu\nu} &=& 8\pi G_* T^*_{\mu\nu} \nonumber\\
       &+&  2\partial_\mu\varphi_*\partial_\nu\varphi_*
          - g^*_{\mu\nu}\left(\partial_\alpha\varphi_*\right)^2
          - 2g^*_{\mu\nu}V\label{einframeeintein}\\
 (\nabla_\mu\nabla^\mu)_*\varphi_* &=& V_{,\varphi_*} - 4\pi G_* \alpha(\varphi_*)
            T^*_{\mu\nu}g_*^{\mu\nu}\label{ekgEF}\\
 \nabla_\mu T^{\mu\nu}_* &=& \alpha(\varphi_*)
                 T^*_{\sigma\rho}g_*^{\sigma\rho} \partial^\nu\varphi_*
\end{eqnarray}
where we have defined the Einstein frame stress-energy tensor
$$
 T^{\mu\nu}_* \equiv \frac{2}{\sqrt{-g_*}}\frac{\delta S_{\rm matter}}{\delta g^*_{\mu\nu}},
$$
related to the Jordan frame stress-energy tensor by
$T_{\mu\nu}^*=A^2T_{\mu\nu}$. The function
\begin{equation}\label{eqalpha}
 \alpha(\varphi_*)\equiv \frac{\dd\ln A}{\dd\varphi_*}.
\end{equation}
characterizes the coupling of the scalar field to matter (we
recover general relativity with a minimally coupled scalar field
when it vanishes). For completeness, we also introduce
\begin{equation}\label{eqbeta}
 \beta(\varphi_*)\equiv \frac{\dd\alpha}{\dd\varphi_*}.
\end{equation}
Note that in Einstein frame the Einstein equations
(\ref{einframeeintein}) are the same as those obtained in general
relativity with a minimally coupled scalar field.

The action~(\ref{actionJF}) defines an effective gravitational
constant $G_\eff = G_*/F = G_*A^2$. This constant does not
correspond to the gravitational constant effectively measured in a
Cavendish experiment. The Newton constant measured in this
experiment is
\begin{equation}
 G_\cav = G_*A_0^2(1+\alpha_0^2)
\end{equation}
where the first term, $G_*A_0^2$ corresponds to the exchange of a
graviton while the second term $G_*A_0^2\alpha_0^2$ is related to
the long range scalar force.

\subsubsection{Cosmological signatures}

The post-Newtonian parameters can be expressed in terms of the
values of $\alpha$ and $\beta$ today as
\begin{equation}
 \gamma^\ppn - 1 = -\frac{2\alpha_0^2}{1+\alpha^2_0},\qquad
 \beta^\ppn - 1 =\frac{1}{2}
 \frac{\beta_0\alpha_0^2}{(1+\alpha_0^2)^2}.
\end{equation}
The Solar system constraints discussed in \S~\ref{sec1222} imply
$\alpha_0$ to be very small, typically $\alpha_0^2<10^{-5}$ while
$\beta_0$ can still be large. Binary pulsar
observations~(Esposito-Far\`ese, 2005) impose that $\beta_0>-4.5$.

\begin{figure}
    \centering
    \includegraphics[width=7cm]{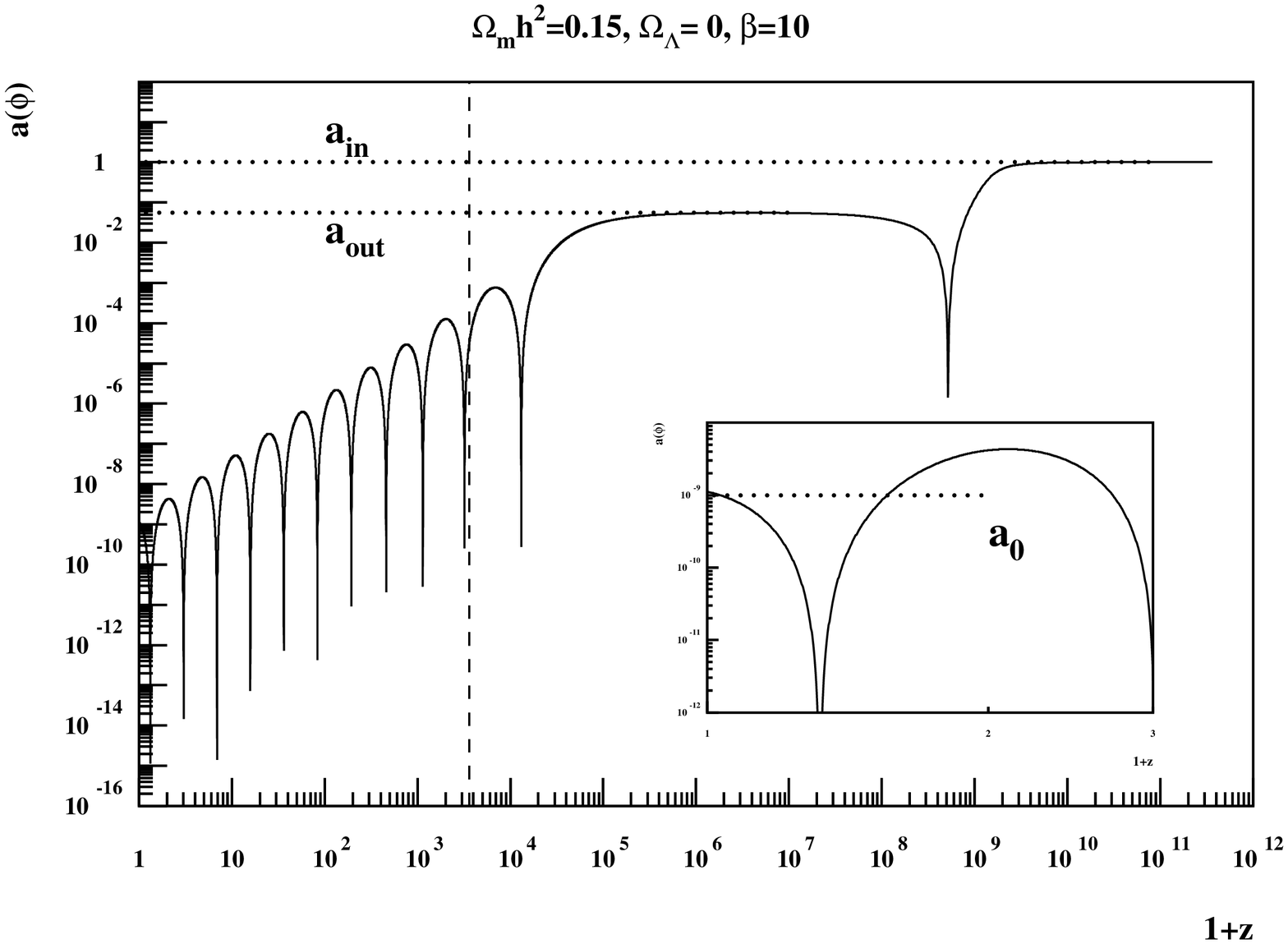}
    \includegraphics[width=5.2cm]{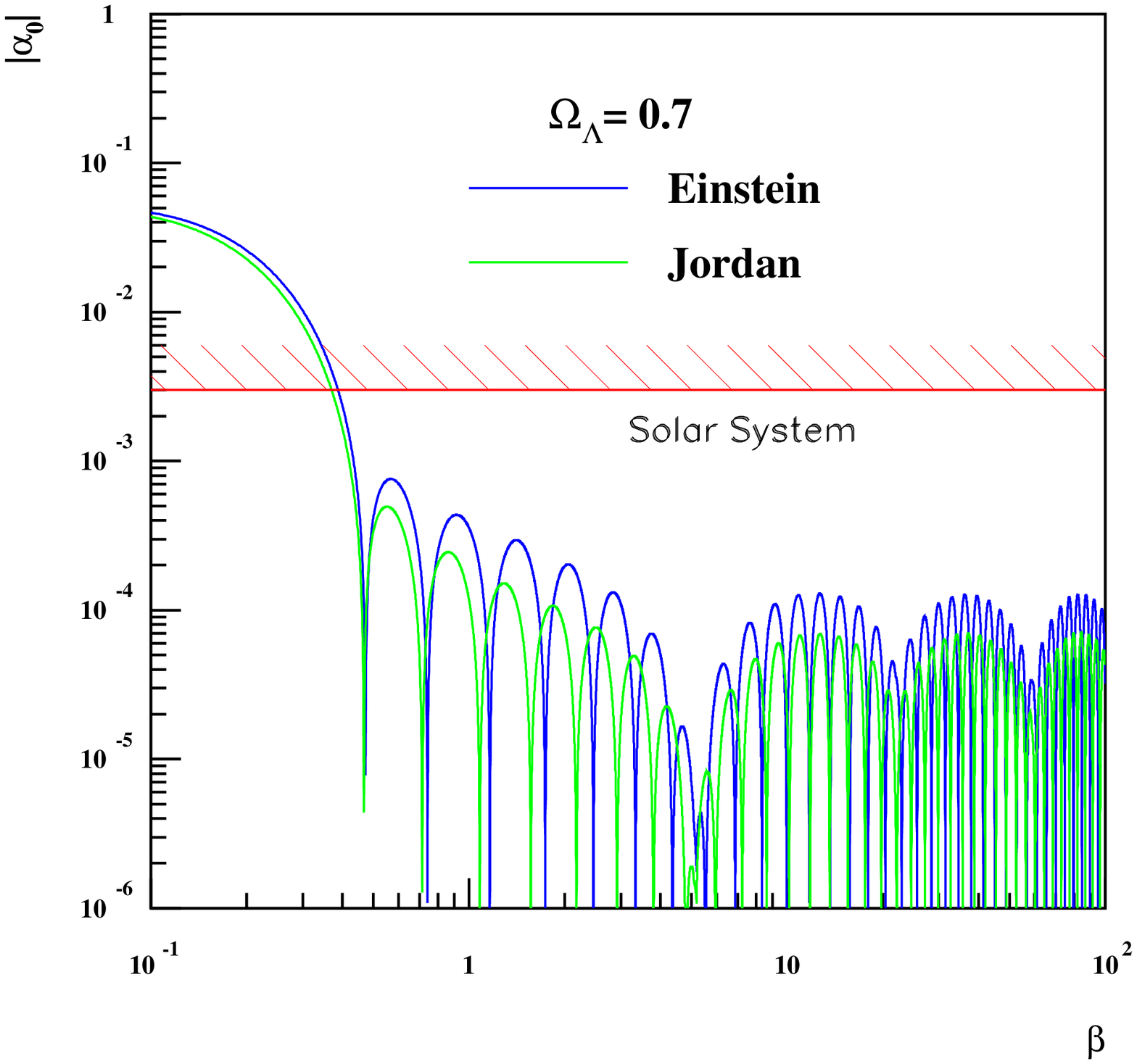}
    \caption{{\it Left}: Evolution of the dilaton as a function of
    redshift. In the radiation era the dilaton freezes to a
    constant value and is then driven toward the minimum of the
    coupling function during the matter era.
    {\it Right}: constraints on scalar-tensor theories of gravity with a
    massless
    dilaton with quadratic coupling in the $(\alpha_0,\beta)$ plane. At large $\beta$
    the primordial nucleosynthesis sets more stringent constraints than the Solar system.
    From Coc \etal (2006).}
    \label{fig1}
  \end{figure}

The previous constraints can be satisfied even if the
scalar-tensor theory was far from general relativity in the past.
The reason is that these theories can be attracted toward general
relativity~(Damour and Nordtvedt, 1993) if their coupling function
or potential has a minimum. This can be illustrated in the case of
a massless ($V=0$) dilaton with quadratic coupling ($a\equiv\ln A=
\frac12\beta\varphi_*^2$). The Klein-Gordon equation~(\ref{ekgEF})
can be rewritten in terms of the number of e-folds in Einstein
frame as
\begin{eqnarray}\label{kgqq}
 \frac{2}{3-\varphi_*^{'2}}\varphi_*''
 +(1-w)\varphi_*' =-\alpha(\varphi_*)(1-3w).
\end{eqnarray}
As emphasized by Damour and Nordtvedt (1993), this is the equation
of motion of a point particle with a velocity dependent inertial
mass, $m(\varphi_*)=2/(3-\varphi_*^{'2})$ evolving in a potential
$\alpha(\varphi_*)(1-3w)$ and subject to a damping force,
$-(1-w)\varphi_*'$. During the cosmological evolution the field is
driven toward the minimum of the coupling function. If $\beta>0$,
it drives $\varphi_*$ toward 0, that is $\alpha\rightarrow0$, so
that the scalar-tensor theory becomes closer and closer to general
relativity. When $\beta<0$, the theory is driven away from general
relativity and is likely to be incompatible with local tests
unless $\varphi_*$ was initially arbitrarily close to 0.

During the radiation era, $w=\frac13$ and the coupling is not
efficient so that $\varphi_*$ freezes to a constant value. Then,
during the matter era, the coupling acts as a potential with a
minimum in zero, hence driving $\varphi_*$ towards zero and the
theory towards general relativity (see Fig.~\ref{fig1}).

This offers a rich phenomenology for cosmology and in particular
for the dark energy question. It was shown that quintessence
models can be extended to scalar-tensor theory of gravity~(Uzan,
1999; Bartolo and Pietroni, 2000) and that it offers the
possibility to have an equation of state smaller than $-1$ with a
well-defined theory~(Martin \etal, 2006). The constraints on the
deviations from general relativity can also be sharpened by the
use of cosmological observations such as cosmic microwave
background anisotropies~(Riazuelo and Uzan, 2002), weak
gravitational lensing~(Schimd \etal, 2005) and big-bang
nucleosynthesis~(Coc \etal, 2006). Fig.~\ref{fig1} summarizes the
constraints that can be obtained from primordial nucleosynthesis.

\subsubsection{Note on $f(R)$ models}\label{secfR}

As discussed in~\S~\ref{sechog}, the only higher order
modifications of the Einstein-Hilbert leading to a well-defined
theory are
\begin{eqnarray}\label{11.NLaction}
  S =\frac{1}{16\pi G_*}\int f(R)\sqrt{-g}\dd^4 x
      + S_{\rm matter}[g_{\mu\nu};\mathrm{matter}].
\end{eqnarray}
Such a theory leads to the field equations
\begin{equation}\label{11.NLeisntein}
 f'(R)R_{\mu\nu} -\frac{1}{2}f(R)g_{\mu\nu}
 -\nabla_\mu\partial_\nu f'(R) + g_{\mu\nu}\Box f'(R) = 8\pi G_*
 T_{\mu\nu},
\end{equation}
where a prime indicates a derivative of the function with respect
to its argument, i.e. $f'(R)\equiv \dd f/\dd R$.

Interestingly, one can show that these theories reduce to a
scalar-tensor theory~(Gottl\"ober \etal, 1990; Teyssandier and
Tourrenc, 1993; Mangano and Sokolowski, 1994; Wands, 1994). To
show this, let us introduce an auxiliary field $\varphi$ and
consider the action
\begin{eqnarray}\label{11.NLaction2}
  S &&=\frac{1}{16\pi G_*}\int \left[f'(\varphi)R + f(\varphi) - \varphi
  f'(\varphi)\right] \sqrt{-g}\dd^4 x\nonumber \\
  &&\qquad +
  S_{\rm matter}[g_{\mu\nu};\mathrm{matter}].
\end{eqnarray}
The variation of this action with respect to the scalar field
indeed implies, if  $f''(\varphi)\not=0$ (The case $f''=0$ is
equivalent to general relativity with a cosmological constant),
that
\begin{equation}
 R-\varphi = 0.
\end{equation}
This constraint permits to rewrite Eq.~(\ref{11.NLeisntein}) in
the form
\begin{equation}\label{11.NLeinstein2}
 f'(\varphi)G_{\mu\nu} -\nabla_\mu\partial_\nu f'(\varphi)
+ g_{\mu\nu}\Box f'(\varphi)
 +\frac{1}{2}[\varphi f'(\varphi)-f(\varphi)]g_{\mu\nu}
   = 8\pi G_* T_{\mu\nu},
\end{equation}
which then reduces to Eq.~(\ref{einframeeintein}) after the field
redeifinitions necessary to shift to the Jordan frame. Note that,
even if the action~(\ref{11.NLaction2}) does not possess a kinetic
term for the scalar field, the theory is well defined since the
true spin-0 degree of freedom clearly appears in the Einstein
frame, and with a positive energy.

The change of variable~(\ref{jf_to_ef1}) implies that we can
choose $\varphi_* = \frac{\sqrt{3}}{2}\ln f'(\varphi)$ so that the
theory in the Einstein frame is defined by
\begin{equation}
 A^2 \propto \hbox{e}^{-\frac{4\varphi_*}{\sqrt{3}}},  \qquad
 V = \frac{1}{4}\left\lbrace
 \varphi(\varphi_*)\hbox{e}^{\frac{2\varphi_*}{\sqrt{3}}}
   -f[\varphi(\varphi_*)]\right\rbrace\hbox{e}^{-\frac{4\varphi_*}{\sqrt{3}}}.
\end{equation}
Note that $\alpha_0$ cannot be made arbitrarily small since the
form of the coupling function $A$ arises from the function $f$. In
order to make these models compatible with Solar system
constraints, the potential should be such that the scalar field is
massive enough, while still being bounded from below.

This example highlights the importance of looking for the true
degrees of freedom of the theory. A field redefinition can be a
useful tool to show that two theories are actually equivalent.
This result was generalized~(Wands, 1994) to theories involving
$f[R,\Box R,\ldots,\Box^nR]$ which were shown to be equivalent to
$(n+1)$ scalar-tensor theories.

This equivalence between $f(R)$ and scalar-tensor theories assumes
that the Ricci scalar is a function of the metric and its first
derivatives. There is a difference when one considers $f(R)$
theories in the Palatini formalism~(Flanagan, 2004), in which the
metric and the connections are assumed to be independent fields,
since while still being equivalent to scalar-tensor theories, the
scalar field does not propagate because it has no kinetic term in
the Einstein frame. It thus reduces to a Lagrange parameter whose
field equation sets a constraint.

\subsubsection{Extensions}

The previous set-up can easily be extended to include $n$ scalar
fields~(Damour and Esposito-Far\`ese, 1992) in which case the
kinetic term will contain a $n\times n$ symmetric matrix,
$g^{\mu\nu}_*\gamma_{ab}(\varphi_c)
\partial_\mu\varphi^a\partial_\nu\varphi^b$.

Another class of models arises when one considers more general
kinetic terms of the form $f(s,\varphi)$ where $s=g^{\mu\nu}_*
\partial_\mu\varphi\partial_\nu\varphi$. When the coupling function reduces
to $A=1$, these models are known as K-essence~(Armendariz-Picon
\etal, 2000; Chiba \etal, 2000). We refer to Esposito-Far\`ese and
Bruneton (2007) and Bruneton (2006) for a discussion of the
conditions to be imposed on $f$ in order for such a theory to be
well-defined.

\subsubsection{Reconstructing theories}

This section has illustrated the difficulty of modifying
consistently general relativity. Let us emphasize that most of the
models we discussed contain several free functions and general
relativity in some continuous limit. It is clear that most of them
cannot be excluded observationally.

It is important to remember that we hope these theories to go
beyond a pure description of the data. In particular, it is
obvious that the function $E(z)$ defined in Eq.~(\ref{defEz}) for
a $\Lambda$CDM model can be reproduced by many different models.
In particular, one can always design a scalar field model inducing
an energy density $\rho_\de(z)$, obtained from the observed
function $H^2(z)$ by subtracting the contributions of the matter
we know (i.e. pressureless matter and radiation). Its potential is
given by~(Uzan, 2007)
\begin{eqnarray}
 V(a) &=& \frac{H(1-X)}{16\pi G}\left(6H+2aH'-\frac{aHX'}{1-X}
 \right)\ ,
 \nonumber\\
 Q(a) &=& \int \frac{\dd\ln a}{\sqrt{8\pi G}}\left[aX'-2(1-X)a\frac{H'}{H}
 \right]\ ,
\end{eqnarray}
with $X(a) \equiv 8\pi G\rho_\de(a)/3H^2(a)$ in order to
reproduce$\lbrace H(a),\rho_\de(a)\rbrace$.

The background dynamics provides only one observable function,
namely $H(z)$, so that it can be reproduced by many theories
having at least one free function.  To go further, we must add
independent information, which can be provided e.g. by the growth
rate of the large scale structure. An illustrative game was
presented in Uzan (2007) in which it was shown that while the
background dynamics of the DGP model (Dvali \etal, 2000) can be
reproduced by a quintessence model, both models did not share the
same growth rate and can be distinguished, in principle, at this
level. However, both the background and sub-Hubble perturbation
dynamics of the DGP model can be reproduced by a well-defined
scalar-tensor theory, which has two arbitrary functions. The only
way to distinguish the two models is then to add local information
since the scalar-tensor theory that reproduces the cosmological
dynamics of the DGP model would induce a time variation of the
gravitational constant above acceptable experimental limits.

This shows the limit of the model-dependent approach in which a
reconstructed theory could simply be seen as a description of a
set of data if its number of free functions is larger than the
observable relations provided by the data. The reconstruction
method can however lead to interesting conclusions and to the
construction of counter-examples. For instance, it was
shown~(Esposito-Far\`ese and Polarski, 2001) that a scalar-tensor
theory with $V=0$ cannot reproduce the background dynamics of the
$\Lambda$CDM.

This should encourage us to consider the simplest possible
extension, namely with the minimum number of new degrees of
freedom and arbitrary functions. In that sense the $\Lambda$CDM
model is very economical since it reproduces all observations at
the expense of a single new constant.

\subsection{Beyond the Copernican principle}\label{sec125}

As explained above, the conclusion that the cosmic expansion is
accelerated is deeply related to the Copernican principle. Without
such a uniformity principle, the reconstruction of the geometry of
our spacetime becomes much more involved.

\begin{figure}
    \centering
    \includegraphics[width=10cm]{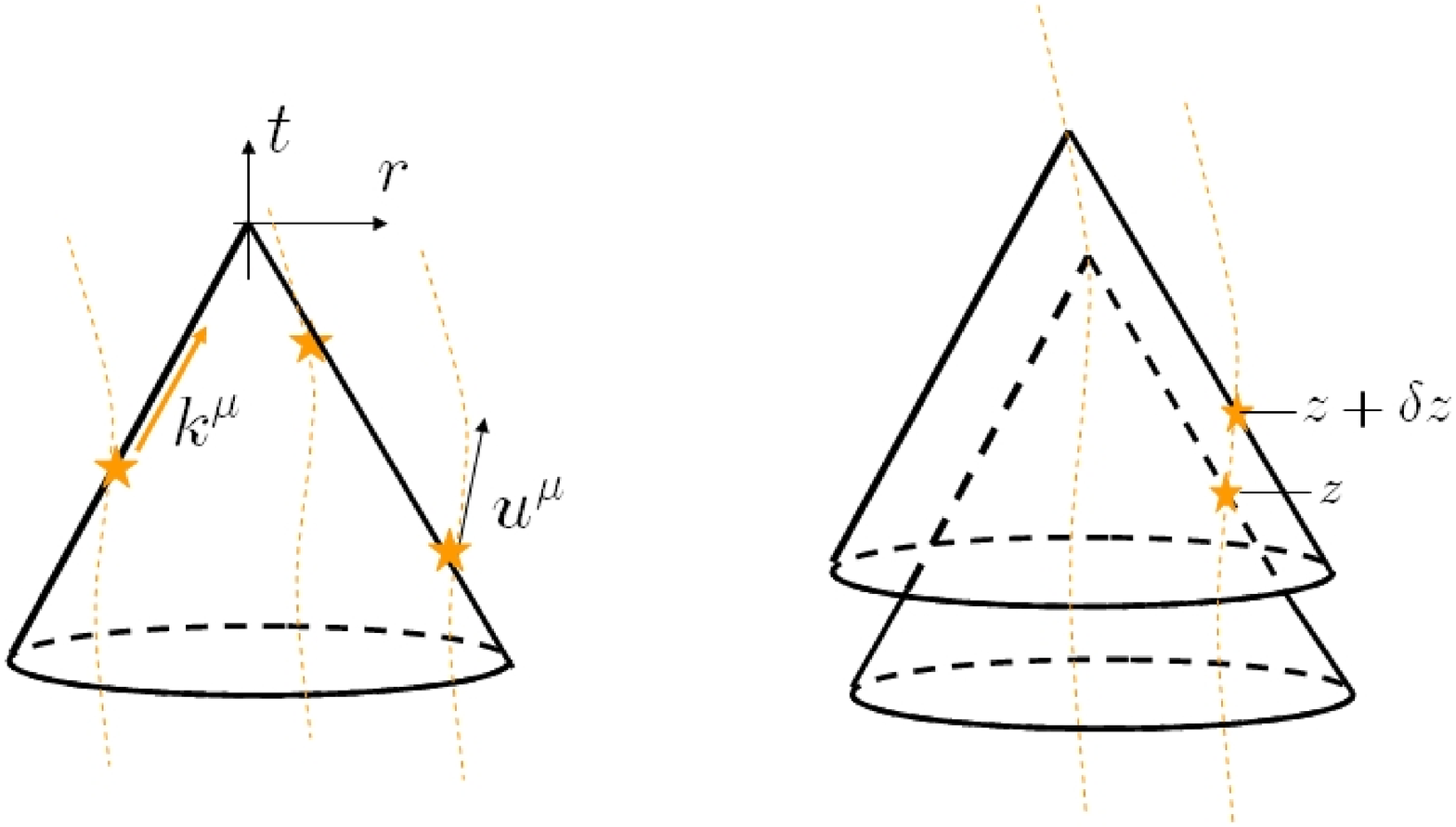}
    \caption{{\it Left}: Most low-redshift data are localized on our past light-cone.
    In a non-homogeneous spacetime there is no direct relation between
    the redshift that is observed and the cosmic time, needed to reconstruct
    the expansion history.
    {\it Right}: The time drift of the redshift allows to extract information
    about two infinitely close past light-cones. $\delta z$ depends
    on the proper motions of the observer and the sources as well
    as the spacetime geometry.}
    \label{fig5}
  \end{figure}

Indeed, most low redshift observations provide the measurements of
some physical quantities (luminosity, size, shape\ldots) as a
function of the position on the celestial sphere and the redshift.
In any spacetime, the redshift is defined as
\begin{equation}
 1+z = \frac{\left(u^\mu k_\mu\right)_{\rm emission}}{\left(
        u^\mu k_\mu\right)_{\rm observation}},
\end{equation}
where $u^\mu$ is the 4-velocity of the cosmic fluid and $k^\mu$
the tangent vector to the null geodesic relating the emission and
the observation (see Fig.~\ref{fig5}). The redshift depends on the
structure of the past light-cone and thus on the symmetries of the
spacetime. It reduces to the simple expression~(\ref{z.flrw}) only
for a Roberston-Walker spacetime. Indeed, it is almost impossible
to prove that a given observational relation, such as the
magnitude-redshift relation, is not compatible with an other
spacetime geometry.

While isotropy around us seems well established observationally
(see e.g. Ruiz-Lapuente, 2007), homogeneity is more difficult to
test. The possibility, that we may be living close to the center
of a large under-dense region has sparked considerable interest,
because such models can successfully match the magnitude-redshift
relation of type Ia supernovae without the need to modify general
relativity or add dark energy.

In particular, the low redshift (background) observations such as
the magnitude-redshift relation can be matched~(C\'el\'erier,
2000; Tomita, 2001; Iguchi \etal, 2002; Ellis, 2008) by a
non-homo\-geneous spacetime of the Lema\^{\i}tre-Tolman-Bondi
(LTB) family, i.e. a spherically symmetric solution of Einstein
equations sourced by pressureless matter and no cosmological
constant.

The geometry of a LTB spacetime~(Lema\^{\i}tre, 1933; Tolman,
1934; Bondi, 1947) is described by the metric
$$
 \dd s^2 =-\dd t^2 + S^2(r,t)\dd r^2 + R^2(r,t)\dd\Omega^2
$$
where $S(r,t)= R'/\sqrt{1+2E(r)}$ and $\dot
R^2=2M(r)/R(r,t)+2E(r)$, using a dot and prime to refer to
derivatives with respect to $t$ and $r$ respectively. The Einstein
equations can be solved parametrically as
\begin{equation}
 \{R(r,\eta),t(r,\eta)\}=\left\{\frac{M(r)}{{\cal
 E}(r)}\Phi'(\eta),T_0(r)+\frac{M(r)}{[{\cal E}(r)]^{3/2}}\,\Phi(\eta) \right\}
\end{equation}
where $\Phi$ is defined by $\Phi(\eta)=\left(\sinh\eta
-\eta,\eta^3/6,\eta-\sin\eta\right)$, and ${\cal E}(r)= (2E,$
$2,-2E)$ according to whether $E$ is positive, null or negative.

This solution depends on 3 arbitrary functions of $r$ only,
$E(r)$, $M(r)$ and $T_0(r)$. Their choice determines the model
completely. For instance $(E,M,T_0)=(-K_0r^2,M_0r^3,0)$
corresponds to a Robertson-Walker spacetime. One can further use
the freedom in the choice of the radial coordinate to fix one of
the three functions at will so that one effectively has only 2
arbitrary independent functions.

Let us sketch the reconstruction and use $r$ as the integration
coordinate, instead of $z$. Our past light-cone is defined as
$t=\hat t(r)$ and we set ${\cal R}(r)\equiv R[\hat t(r),r]$. The
time derivative of $R$ is given by $\dot R[\hat t(r),r]\equiv{\cal
R}_1=\sqrt{2M_0r^3/{\cal R}(r) +2E(r)}$. Then we get $R'[\hat
t(r),r]\equiv{\cal R}_2(r) = -[{\cal R}(r) -3(\hat t(r)
-T_0(r)){\cal R}_1(r)/2]E'/E - {\cal R}_1(r)T_0'(r) +{\cal
R}(r)/r$. Finally, more algebra leads to $\dot R'[\hat
t(r),r]\equiv{\cal R}_3(r)=[{\cal R}_1(r) -3 M_0r^3(\hat t(r)
-T_0(r))/{\cal R}^2(r)]E'(r)/2E(r)+M_0r^3T_0'(r)/{\cal R}^2 +{\cal
R}_1(r)/r$. Thus, $\dot R$, $R'$ and $\dot R'$ evaluated on the
light cone are just functions of ${\cal R}(r)$, $E(r)$, $T_0(r)$
and their first derivatives. Now, the null geodesic equation gives
that
$$
 \frac{\dd\hat t}{\dd r}= -\frac{{\cal
 R}_2(r)}{\sqrt{1+2E(r)}},\quad
 \frac{\dd z}{\dd r}=\frac{1+z}{\sqrt{1+2E(r)}}{\cal R}_3(r),
$$
and
$$
\frac{\dd {\cal R}}{\dd r}=\left[ 1 - \frac{{\cal
R}_1(r)}{\sqrt{1+2E(r)}}\right]
 {\cal R}_2(r).
$$
These are 3 first order differential equations relating 5
functions ${\cal R}(r)$, $\hat t(r)$, $z(r)$ $E(r)$ and $T_0(r)$.
To reconstruct the free functions we thus need 2 observational
relations. The reconstruction from background data alone is
under-determined and one must fix one function by hand. The
angular distance-redshift relation, ${\cal R}(z)=D_A(z)$, is the
obvious choice. This explains why the magnitude-redshift relation
can be matched~(C\'el\'erier, 2000; Tomita, 2001; Iguchi \etal,
2002) by a LTB geometry? Indeed the geometry is not fully
reconstructed.

It follows that many issues are left open. First, can we use more
observational data to close the reconstruction of the LTB
geometry. Indeed the knowledge of the growth rate of the large
scale structure could be used, as for the reconstruction of the
two arbitrary functions of a scalar-tensor theory but no full
investigation of the perturbation theory around a LTB spacetime
has been performed (Zibin, 2008; Dunsby and Uzan, 2009). Second,
can we construct the model-independent test of the Copernican
principle avoiding the necessity to restrict to a given geometry
since we may have to consider more complex spacetimes that the LTB
one. Third, we would have to understand how these models reproduce
the predictions of the standard cosmological model on large scales
and at early times, e.g. how are the cosmic microwave background
anisotropies and the big-bang nucleosynthesis dependent on these
spacetime structures.

\subsection{Conclusions}

This section has investigated two different ways to modify our
reference cosmological model by either extending the description
of the laws of nature or by extending the complexity of the
geometry of our spacetime by relaxing the Copernican principle.

Whatever the choice, we see that many possibilities are left open.
All of them introduce new degrees of freedom, either as physical
fields or new geometrical freedom, and free functions. They also
contain the standard $\Lambda$CDM as a continuous limit (e.g. the
potential can become flat, the arbitrary functions of a LTB can
reduce to their FLRW form etc.) These extensions are thus almost
non-excludable by cosmological observations alone and as we have
seen, they can reduce to pure descriptions of the data. Again, we
must be guided by some principles.

The advantages of the model-dependent approaches is that we  know
whe\-ther we are dealing with well-defined theories or spacetime
structures. All cosmological observables can be consistently
computed so that these models can be safely compared to
observations to quantify how close from a pure $\Lambda$CDM the
model of our Universe should be. They can also forecast the
ability of coming surveys to constrain them.

The drawback is that we cannot test all the possibilities which
are too numerous. An alternative is to design parameterizations
which have the advantage, we hope, to encompass many models. The
problem is then the physical interpretation of the new parameters
that are measured from the observations.

Another route, that we shall now investigate, is to design null
tests of the $\Lambda$CDM model in order to indicate what kind of
modifications, if any, are required by the observations.

\section{Testing the underlying hypotheses}

Let us first clarify what we mean by a {\it null test}. Once the
physical theory and the properties of its cosmological solution
have been fixed, there exist rigidities between different
observable quantities. They reflect the set of assumptions of our
reference cosmological models. By testing these rigidities we can
strengthen our confidence in the principles on which our model
lies. In case we can prove that some of them are violated, it will
just give us a hint in the way to extend our cosmological model
and on which principle has to be questioned.

Let us take a few examples that will be developed below.
\begin{itemize}
 \item The equation of state of the dark energy must be $w_\de=-1$
 and constant in time.
 \item The luminosity and angular distances must be related by the
 distance duality relation stating that $D_L(z)=(1+z)^2D_A(z)$.
 \item On sub-Hubble scales, the gravitational potential and the
 perturbation of the matter energy density must be related by the
 Poisson equation, $\Delta\Phi = 4\pi G\rho_\mat a^2 \delta_\mat$,
 which derives from the Einstein equation in the weak field limit.
 \item On sub-Hubble scales, the background dynamics and the
 growth of structure are not independent.
 \item The constants of nature must be strictly constant.
\end{itemize}
These rigidities are related to different hypotheses, such as the
validity of general relativity or Maxwell theory. We shall now
describe them and see how they can be implemented with
cosmological data.

\subsection{Testing the Copernican principle}

The main difficulty in testing the Copernican principle, as
discussed in \S~\ref{sec125}, lies in the fact that all
observations are located on our past light-cone and that many
four-dimensional spacetimes may be compatible with the same
three-dimensional light-like slice~(Ellis, 1975).

Recently, it was realized that cosmological observations may
however provide a test of the Copernican principle~(Uzan \etal,
2008b). This test exploits the time drift of the redshift that
occurs in any expanding spacetime, as first pointed out in the
particular case of Robertson-Walker spacetimes for which it takes
the form~(Sandage, 1962; McVittie, 1962)
\begin{equation}\label{sandloeb}
 \dot z = (1+z)H_0 - H(z)\ .
\end{equation}
Such an observation gives informations on the dynamics outside the
past light-cone since it compares the redshift of a given source
at two times and thus on two infinitely close past light-cones
(see Fig.~\ref{fig5}-right). It follows that it contains an
information about the spacetime structure along the worldlines of
the observed sources that must be compatible with the one derived
from the data along the past light-cone.

For instance, in a spherically symmetric spacetime, the
expression~(\ref{sandloeb}) depends on the shear, $\sigma(z)$, of
the congruence of the wordlines of the comoving observers
evaluated along our past light-cone,
$$
 \dot z = (1+z)H_0 - H(z) -\frac{1}{\sqrt{3}}\sigma(z)\ .
$$
It follows that, when combined with other distance data, it allows
to determine the shear on our past light-cone and we can check
whether it is compatible with zero, as expected for any
Robertson-Walker spacetime.

In a RW universe, we can go further and determine a consistency
relation between several observables. From the
metric~(\ref{flrw}), one deduces that $H^{-1}(z)= D'(z)\left[ 1 +
\Omega_{K0}H_0^2D^2(z) \right]^{-1/2}$, where a prime stands for
$\partial_z$ and $D(z)=D_L(z)/(1+z)$; this relation being
independent of the Friedmann equations. It follows that in any
Robertson-Walker spacetime the {\it consistency relation},
$$
 1 + \Omega_{K0}H_0^2\left(\frac{D_L(z)}{1+z}\right)^2 -
 \left[H_0(1+z) -\dot z(z)
 \right]^2\left[\frac{\dd}{\dd
 z}\left(\frac{D_L(z)}{1+z}\right)\right]^2=0,
$$
between observables must hold whatever the matter content and the
field equations, since it derives from pure kinematical relations
that do not rely on the dynamics (a similar analysis is provided in
Clarkson \etal, 2008). The measurement of $\dot z(z)$
will also allow~(Uzan \etal, 2008b) to close the reconstruction of
the local geometry of such an under-dense region (as discussed in
\S~\ref{sec125}).

$\dot z(z)$ has a typical amplitude of order $\delta z\sim
-5\times10^{-10}$ on a time scale of $\delta t= 10$~yr, for  a
source at redshift $z=4$. This measurement is challenging, and
impossible with present-day facilities. However, it was recently
revisited in the context of Extremely Large Telescopes (ELT),
arguing they could measure velocity shifts of order $\delta v\sim
1-10\ {\rm cm/s}$ over a 10 years period from the observation of
the Lyman-$\alpha$ forest. It is one of the science drivers in
design of the CODEX spectrograph~(Pasquini \etal, 2005) for the
future European ELT. Indeed, many effects, such as proper motion
of the sources, local gravitational potential, or acceleration of
the Sun may contribute to the time drift of the redshift. It was
shown~(Liske \etal, 2008; Uzan \etal, 2008), however, that these
contributions can be brought to a 0.1\% level so that the
cosmological redshift is actually measured.

Let us also stress that another idea was also recently
proposed~(Goodman, 1995; Caldwell and Stebbins, 2008). It is based
on the distortion of the Planck spectrum of the cosmic microwave
background.

\subsection{Testing General relativity on astrophysical scales}

\subsubsection{Test of local position invariance}

The local position invariance is one aspect of the Einstein
equivalence principle which is at the basis of the hypothesis of
metric coupling. It implies that all constants of nature must be
strictly constant. The indication that the numerical value of any
constant has drifted during the cosmological evolution would be a
sign in favor of models of the classes C and D.

\begin{figure}
    \centering
    \includegraphics[width=10cm]{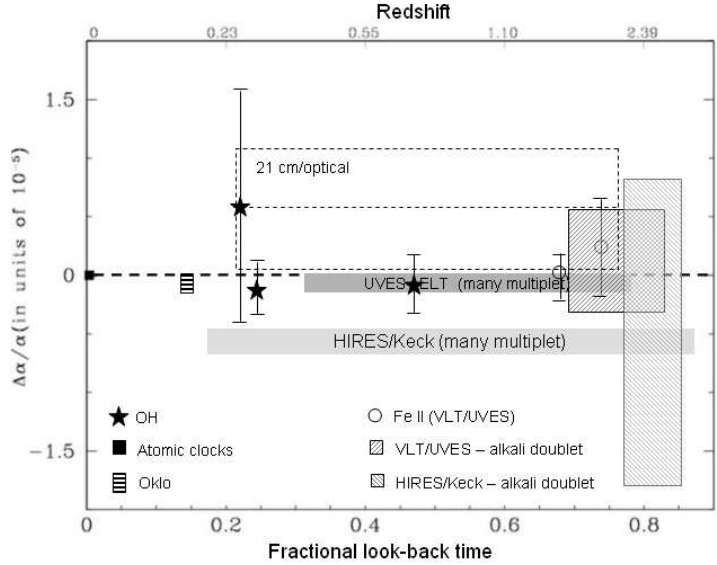}
    \caption{Constraints on the time variation of the fine structure
    constant $\alpha$ from the observations of quasar absorption spectra.}
    \label{fig6}
  \end{figure}

The test of the constancy of the fundamental constants has seen a
very intense activity in the past decade. In particular the
observations from quasar absorption spectra have relaunched a
debate on the possible variation of the fine structure constant.
Recently it was also argued~(Coc \etal, 2007) that a time
variation of the Yukawa couplings may allow to solve the lithium-7
problem that, at the moment, has no other physical explanation.

Constraints can be obtained from many physical systems such as
atomic clocks ($z=0$), the Oklo phenomenon ($z\sim0.14$), the
lifetime of unstable nuclei and meteorite data ($z\sim0.2$),
quasar absorption spectra ($z=0.2-3$), cosmic microwave background
($z\sim10^3$) and primordial nucleosynthesis ($z\sim10^8$). The
time variation of fundamental constants is also deeply related to
the universality of free fall. We refer to Uzan (2003) and (2004)
for extensive reviews on the methods and the constraints, which
are summarized on Figure~\ref{fig6}.

In conclusion, we have no compelling evidence for any time
variation of a constant, which sets strong constraints on the
couplings between the dark energy degrees of freedom and ordinary
matter. We can conclude the local position invariance holds in our
observable universe and that metric couplings are favored.

\subsubsection{Test of the Poisson equation}

Extracting constraints on deviations from GR is difficult because
large scale structures entangle the properties of matter and
gravity. On sub-Hubble scales, one can, however, construct tests
reproducing those in the Solar system. For instance, light
deflection is a test of GR because we can measure independently
the deflection angle and the mass of the Sun.

On sub-Hubble scales, relevant for the study of the large-scale
structure, the Einstein equations reduce to the Poisson equation
\begin{equation}\label{Peq}
\Delta\Psi = 4\pi G\rho_\mat a^2
\delta_\mat=\frac{3}{2}\Omega_\mat H^2 a^2\delta_\mat,
\end{equation}
relating the gravitational potential and the matter density
contrast.

As first pointed out by Uzan and Bernardeau (2001), this relation
can be tested on astrophysical scales, since the gravitational
potential and the matter density perturbation can be measured
independently from the use of cosmic shear measurements and galaxy
catalogs. The test was recently implemented with the CFHTLS-weak
lensing data and the SDSS data to conclude that the Poisson
equation holds observationally to about 10~Mpc~(Dor\'e \etal,
2007).

\begin{figure}
    \centering
    \includegraphics[width=8cm]{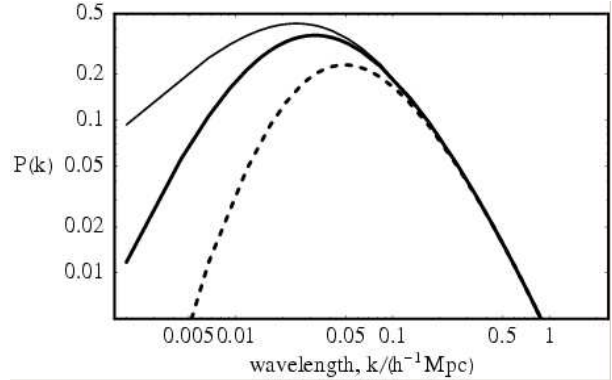}
    \caption{In a theory in which gravity switches from a standard
 four-dimensional gravity to a DGP-like five-dimensional gravity above a
 crossover scale of $r_s=50h^{-1}$~Mpc, there are different cosmological
 implications concerning the growth of cosmological perturbations. Since
 gravity becomes weaker on large scales, fluctuations stop growing.
 It implies that the density contrast power spectrum (thick
 line) differs from the standard one (thin line) but, more important, from
 the gravitational potential power spectrum (dash line). From
 Uzan and Bernardeau (2001).}
    \label{fig8p}
  \end{figure}

As an example, Fig.~\ref{fig8p} depicts the expected modifications
of the matter power spectrum and of the gravitational potential
power spectrum in the case of a theory in which gravity switches
from a standard four-dimensional gravity to a DGP-like
five-dimensional gravity above a crossover scale of
$r_s=50h^{-1}$~Mpc. Since gravity becomes weaker on large scales,
density fluctuations stop growing, exactly as when the
cosmological constant starts dominating. It implies that the
density contrast power spectrum differs from the standard one but,
more important, from the gravitational potential power spectrum.

Let us emphasize that, the deviation from the standard behavior of
the matter power spectrum is model dependent (it depends in
particular on the cosmological parameters), but that the
discrepancy between the matter and gravitational potential
Laplacian power spectra is a direct signature of a modification of
general relativity.

The main limitation in the applicability of this test is due to
the biasing mechanisms (i.e. the fact that galaxies do not
necessarily trace faithfully the matter field) even if it is
thought to have no significant scale dependence at such scales.

\subsubsection{Toward a post-$\Lambda$CDM formalism}

The former test of the Poisson equation exploits one rigidity of
the field equations on sub-Hubble scales. It can be improved by
considering the full set of equations.

Assuming that the metric of spacetime takes the form
\begin{equation}
 \dd s^2 = -(1+2\Phi)\dd t^2 + (1-2\Psi)a^2\gamma_{ij}\dd x^i\dd x^j
\end{equation}
on sub-Hubble scales, the equation of evolution reduces to the
continuity equation
\begin{equation}\label{m1}
 \delta_\mat' + \theta_\mat =0,
\end{equation}
where $\theta$ is the divergence of the velocity perturbation and
a prime denotes a derivative with respect to the conformal time,
the Euler equation
\begin{equation}
 \theta_\mat' +{\cal H}\theta_\mat= -\Delta\Phi,
\end{equation}
where ${\cal H}$ is the comoving Hubble parameter,the Poisson
equation~(\ref{Peq}) and
\begin{equation}\label{m4}
 \Phi=\Psi.
\end{equation}

These equations imply many relation between the cosmological
observables. For instance, decomposing $\delta_\mat$ as
$D(t)\epsilon(x)$ where $\epsilon$ encodes the initial conditions,
the growth rate $D(t)$ evolves as
$$
 \ddot D + 2H\dot D -4\pi G\rho_\mat D = 0.
$$
This equation can be rewritten in terms of $p=\ln a$ as time
variable~(Peter and Uzan, 2005) and considered not as a second
order equation for $D(t)$ but as a first order equation for
$H^2(a)$
$$
 (H^2)'+ 2\left(\frac{3}{a} + \frac{D''}{D'} \right)H^2 =
 3\frac{\Omega_{\mat0}H_0^2 D}{a^2 D'}
$$
where a prime denotes a derivative with respect to $p$. It can be
integrated as~(Chiba and Nakamura, 2007)
\begin{equation}\label{ddd}
 \frac{H^2(z)}{H_0^2} = 3\Omega_{\mat0}\left(\frac{1+z}{D'(z)}\right)^2
 \int\frac{D}{1+z}(-D')\dd z.
\end{equation}
This exhibits a rigidity between the growth function and the
Hubble parameter. In particular the Hubble parameter determined
from background data and from perturbation data using
Eq.~(\ref{ddd}) must agree. This was used in the analysis of Wang
\etal (2007).

Another relation exist between $\theta_\mat$ and $\delta_\mat$.
The Euler equation implies that
\begin{equation}
 \theta_\mat=-\beta(\Omega_{\mat0},\Omega_{\Lambda0})\delta_\mat,
\end{equation}
with
\begin{equation}
 \beta(\Omega_{\mat0},\Omega_{\Lambda0})\equiv \frac{\dd\ln D(a)}{\dd\ln a}.
\end{equation}

We conclude that the perturbation variables are not independent
and the relation between them are inherited from some assumptions
on the dark energy. Phenomenologically, we can generalize
Eqs.~(\ref{m1}-\ref{m4}) to
\begin{eqnarray}
 &&\delta_\mat' + \theta_\mat =0,\\
 &&\theta_\mat' +{\cal H}\theta_\mat= -\Delta\Phi + S_\de,\\
 && -k^2\Phi = 4\pi G F(k,H) \delta_\mat + \Delta_\de,\\
 &&\Delta(\Phi-\Psi) = \pi_\de.
\end{eqnarray}
We assume that there is no production of baryonic matter so that
the continuity equation is left unchanged. $S_\de$ describes the
interaction between dark energy and standard matter. $\Delta_\de$
characterizes the clustering of dark energy, $F$ accounts for a
scale dependence of the gravitational interaction and $\pi_\de$ is
an effective anisotropic stress. It is clear that the $\Lambda$CDM
corresponds to $(F,\pi_\de,\Delta_\de,S_\de) = (1,0,0,0)$. The
expression of $(F,\pi_\de,\Delta_\de,S_\de)$ for quintessence,
scalar-tensor, $f(R)$ and DGP models and more generally for models
of the classes A-D can be found in Uzan (2007).

From an observational point of view, weak lensing survey gives
access to $\Phi+\Psi$, galaxy maps allow to reconstruct
$\delta_g=b\delta_\mat$ where $b$ is the bias, velocity fields
give access to $\theta$. In a $\Lambda$CDM, the correlations
between these observable are not independent since, for instance
$\langle\delta_g\delta_g\rangle = b^2\langle\delta_\mat^2\rangle$,
$\langle\delta_g\theta_m\rangle = -b\beta
\langle\delta_\mat^2\rangle$ and $\langle\delta_g\kappa\rangle =
8\pi G\rho_\mat a^2 b\langle\delta_\mat^2\rangle$.

Various ways of combining these observables have been proposed,
construction of efficient estimators and forecast for possible
future space mission designed to make these tests as well as the
possible limitations (arising e.g. from non-linear bias, the
effect of massive neutrinos or the dependence on the initial
conditions) are now being extensively studied~(Zhang \etal, 2007;
Jain and Zhang, 2007; Amendola \etal, 2008; Song and Koyama,
2008).

To finish let us also mention that the analysis of the weakly
non-linear dynamics allows to develop complementary tests of the
Poisson equation~(Ber\-nardeau, 2004) but no full investigation in
the framework presented here has been performed yet.

\subsection{Other possible tests}

\subsubsection{Distance duality}\label{sec1331}

As long as photons travel along null geodesics and the geodesic
deviation equation holds, the source angular distance, $r_{\rm
s}$, and the observer area distance, $r_{\rm o}$, must be related
by the {\it reciprocity relation}~(Ellis, 1971), $r_{\rm s}^2 =
r_{\rm o}^2 (1+z)^2$ regardless of the metric and  matter content
of the spacetime.

Indeed, the solid angle from the source cannot be measured so that
$r_{\rm s}$ is not an observable quantity. But, it can be shown
that, if the number of photons is conserved, the source angular
distance is related to the luminosity distance, $D_L$, by the
relation $D_L=r_{\rm s}(1+z)$. It follows that there exist a {\it
distance duality relation} between the luminosity and angular
distances,
\begin{equation}\label{reci}
 {D_L}={D_A}(1+z)^2.
\end{equation}
This distance duality relation must hold if the reciprocity
relation is valid and if the number of photons is conserved. In
fact, one can show that in a metric theory of gravitation, if
Maxwell equations are valid, then both the reciprocity relation
and the area law are satisfied and so is the distance duality
relation.

There are many possibilities for one of these conditions to be
violated. For instance the non-conservation of the number of
photons can arise from absorption by dust, but more exotic models
involving photon-axion oscillation in an external magnetic
field~(Csaki \etal, 2002; Deffayet \etal, 2002) (class B) can also
be a source of violation. More drastic violations would arise from
theories in which gravity is not described by a metric theory and
in which photons do not follow null geodesic.

A test of this distance duality relies on the X-ray observations
and Sunyaev-Zel'dovich (SZ) effect of galaxy clusters~(Uzan \etal,
2004).

Galaxy clusters are known as the largest gravitationally bound
systems in the universe. They contain large quantities of hot and
ionized gas which temperatures are typically $10^{7-8}$~K. The
spectral properties of intra-cluster gas show that it radiates
through bremsstrahlung in the X-ray domain. Therefore, this gas
can modify the cosmic microwave background spectral energy
distribution through inverse Compton interaction of photons with
free electrons. This is the so-called SZ effect. It induces a
decrement in the cosmic microwave background brightness at low
frequencies and an increment at high frequencies.

In brief, the method is based on the fact that the cosmic
microwave background temperature (i.e. brightness) decrement due
to the SZ effect is given by $\Delta T_{\rm SZ} \sim
L\overline{n_eT_e}$ where the bar refers to an average over the
line of sight and $L$ is the typical size of the line of sight in
the cluster. $T_e$ is the electron temperature and $n_e$ the
electron density. Besides, the total X-ray surface brightness is
given by $S_X \sim \frac{V}{4\pi D_L^2}\overline{n_en_pT_e^{1/2}}$
where the volume $V$ of the cluster is given in terms of its
angular diameter by $V=D_A^2\theta^2L$. It follows that
\begin{equation}
 S_X \sim \frac{\theta^2}{4\pi}\frac{D_A^2}{D_L^2}L
 \overline{n_en_pT_e^{1/2}}.
\end{equation}
The usual approach is to assume the distance duality relation so
that forming the ratio $\Delta T_{\rm SZ}^2/S_X$ eliminates $n_e$.

\begin{figure}
    \centering
    \includegraphics[width=9cm]{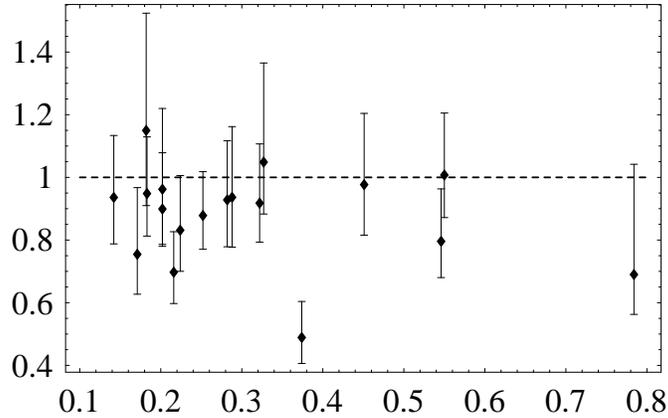}
    \caption{Test of distance duality. The constraint of $\eta$ defined in
    Eq.~(\ref{calc_eta}) at different redshift are obtained by combining
    SZ and X-ray measurements from 18 clusters. No sign of violation of the
    distance duality relation is seen. From Uzan \etal (2004).}
    \label{fig8}
  \end{figure}

We can however use these observation to measure
\begin{equation}\label{calc_eta}
 \eta(z)=\frac{D_L(z)}{(1+z)^2D_A(z)}
\end{equation}
and thus test whether $\eta=1$. Fig.~\ref{fig8} summarizes the
constraints that have been obtained from the analysis of 18
clusters. No sign of violation of the distance duality relation is
seen, contrary to an early claim by Bassett and Kunz (2004).

\subsubsection{Gravity waves}

In models involving two metrics, gravitons and standard matter are
coupled to different metrics. It follows that the propagation of
gravity waves and light may be different. As a consequence the
arrival times of gravity wave and light should not be equal.

An estimation~(Kahya and Woodard, 2007) in the case of the TeVeS
theory for the supernovae 1987A indicates that light shall arrive
days before the gravity waves, which should be easily detectable.

We also emphasize that models in which gravity waves propagate
slower than electromagnetic waves are also very constrained by the
observations of cosmic rays~(Moore and Nelson, 2001) because
particles propagating faster than the gravity waves emit
gravi-Cerenkov radiation.

These two examples highlight that the cosmological tests of
general relativity {\em do not reduce} to the study of the large
scale structures.

\subsubsection{A word on topology}

The debate concerning the topology of our Universe is the
continuation of the historically long discussed question on
whether our Universe is finite or infinite.

The hypothesis on the global topology does no influence local
physics and let most of the theoretical and observational
conclusions unchanged. Local geometry has however a deep impact
since it sets the topologies that are acceptable. In our
cosmological model, the Copernican principle implies that we are
dealing with 3-dimensional spaces of constant curvature. Besides,
almost spatial flatness limits those topologies that would be
detectable~(Weeks \etal, 2003).

A non-trivial topology would violate global isotropy and let
signatures mainly on the statistical isotropy of the cosmic
microwave background ani\-sotropies~(Gaussman \etal, 2001; Luminet
\etal, 2003, Riazuelo \etal, 2004; Riazuelo \etal, 2004b; Uzan
\etal, 2004). The current constraints imply~(Shapiro \etal, 2007)
that the size of the Universe has to be larger than 0.91 times the
diameter of the last scattering surface, that is 24~Gpc.

Even though the cosmological constant can be related to a
characteristic size of the order of $\Lambda^{-1/2}\sim H_0^{-1}$,
no mechanism relating the size of the Universe and the
cosmological constant has been constructed (Calder and Lahav
(2008) however suggests a possible relation to the Mach
principle).

\section{Conclusion}

The acceleration of the cosmic expansion and the understanding of
its origin drives us to reconsider the construction of our
reference cosmological models.

\begin{figure}
    \centering
    \includegraphics[width=9cm]{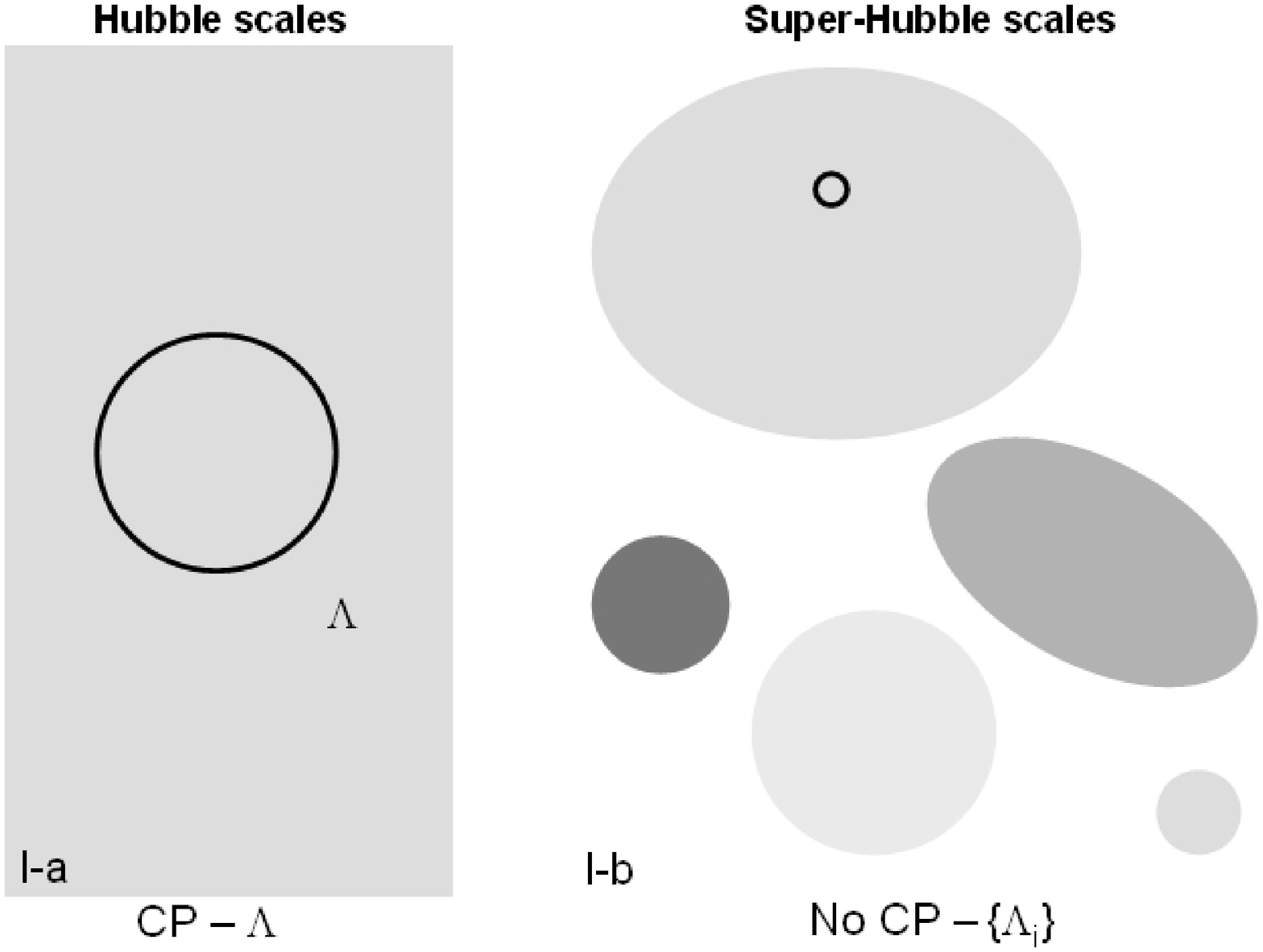}

    \includegraphics[width=9cm]{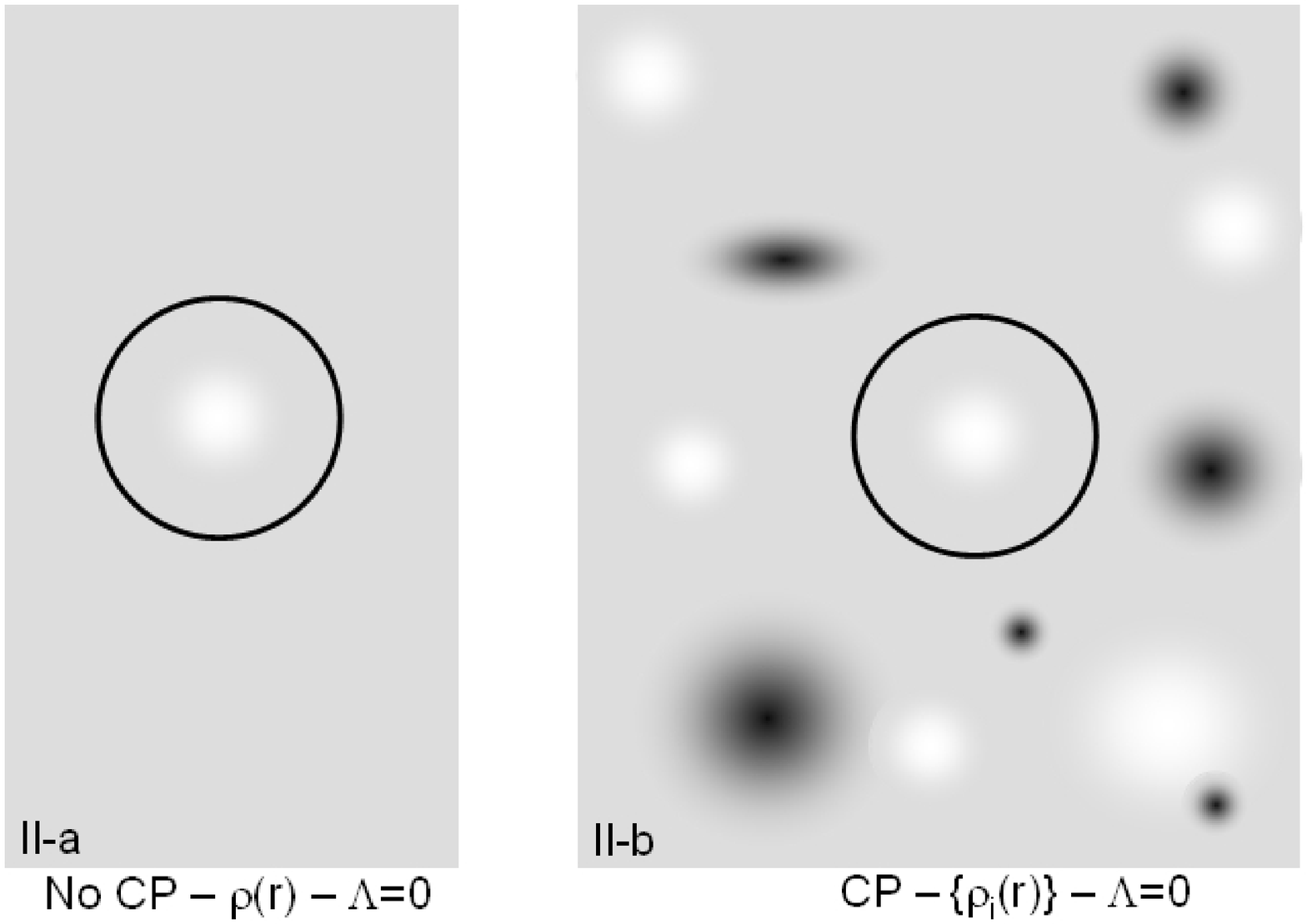}
    \caption{On the scales of the observable universe (circle), the acceleration of the
    universe can be explained by a cosmological constant (or more generally a
    dark energy component) in which case the construction of the cosmological
    model relies on the Copernican principle (upper-left).
    To make sense of a cosmological constant, one introduces a large structure
    known as the multiverse (upper-right) which can be seen as a collection of universes of
    all sizes and in which the values of the cosmological constant, as well as
    other constants, are randomized. The anthropic principle then states that we
    observe only those universes where the value of these constants are such that
    observers can exist. In this sense we have to abandon the Copernican principle
    on the scales of the multiverse.
    An alternative
    is to assume that there is no need for a cosmological constant or
    new physics, in which case we have to abandon the Copernican principle
    and assume e.g. that we are living in an under-dense region (lower-left).
    However, we may recover the Copernican principle on larger
    scales if there exist a distribution of over- and under-dense
    regions of all sizes and densities on super-Hubble scales,
    without the need for a multiverse. In such a view, the
    Copernican principle will be violated on Hubble scale, just
    because we live in such a structure which happens to have a
    size comparable to the one of the observable universe.
    }
    \label{fig9}
  \end{figure}

Three possibilities seem open to us.
\begin{itemize}
 \item {\it Stick to the $\Lambda$CDM}. The model is well-defined,
 does not require to extend the low-energy version of the law of
 nature, and is compatible with all existing data. However, in order
 to make sense of the cosmological constant and avoid the
 cosmological constant problem one needs to invoke a very large
 structure~(Weinberg, 1989; Garriga and Vilenkin, 2004; Carr and Ellis, 2008),
 the multiverse,  a collection of universes in which
 the value of the cosmological
 constant, as well as those of other physical constants, are randomized
 in different regions. Such a structure, while advocated on the basis of the
 string landscape~(Suskind, 2006), has no clear mathematical
 definition~(Ellis \etal, 2004) but it aims at suppressing the contingence
 of our physical models (such as their symmetry groups, value
 of constants,\ldots that, by construction, cannot be explained by these
 models) at the price of an anthropic approach which may
 appear as half-way between pure anthropocentrism, fixing us
 at the center of the universe, and the
 cosmological principle, stating that no place can be favored
 in any way.\\
 In such a situation, it is clear that the Copernican principle
 holds on the size of the observable universe and even much
 beyond. However on the scales of the multiverse, it has
 to be abandoned since, according to this view, we can only
 live in regions of the multiverse where the value of the
 cosmological constant is small enough for observers to exist (see Fig.~\ref{fig9}).

 The alternative would be to better understand the computation of
 the energy density of the vacuum.
 \item {\it Assume $\Lambda=0$} and then
  \begin{itemize}
  \item {\it Assume no new physics.} In such a case, we must
  abandon the Copernican principle on the size of the observable
  universe. This leads us to consider more involved solutions of
  known and established physical theories. Indeed, the main
  objection would be to understand why we shall live in such a
  particular place.\\
  Note however that the Copernican principle can be restored on
  much larger scales (i.e. super-Hubble but without the need to
  invoke a structure like the multiverse). On these scales, one
  can argue that there shall exist a distribution of over- and
  under-dense regions of all sizes and density profiles. In this
  sense, we are just living in one of them, in the same sense
  that stars are more likely to be in galaxies (see Fig.~\ref{fig9})
  and the Copernican principle seems to be violated on Hubble scales, just
    because we live in such a structure which happens to have a
    size comparable to the one of the observable universe.
  \item {\it Invoke new physics}. This an be achieved in numerous
  ways. The main constraint is to construct a well-defined theory.
  In such a case the Copernican principle can hold both on the size of the
  observable universe but also on much larger scales.
  \end{itemize}
\end{itemize}

At the moment, none of these three possibilities is satisfactory,
mainly because it forces us to speculate on scales much beyond
those of the observable universe. A last possibility, that was
alluded to in \S~\ref{reg}, is the possibility that the
acceleration is induced by the backreaction of the large scale
structures but this still needs in depth investigation. We have
argued that future cosmological observations can shed some light
on the way to modify our reference cosmological model and extend
the tests of the fundamental laws of physics, such as general
relativity, as well as some extra-hypotheses such as the
Copernican principle and the topology of space. In this sense, we
follow the most standard physical approach in which any null test
that can be done must be done in order to extend our understanding
of the domain of validity of the description of the physical laws
we are using.

From an observational point of view, demonstrating a violation of
the Copernican principle on the size of the observable universe
will indicate that the second solution is the most likely, but
nothing forces us to accept the associated larger spacetime
described in Fig.~\ref{fig9}. If any of the tests presented here,
or other to be designed, is positive then we will have an
indication that the dark energy is not the cosmological constant
and on the kind of extension required. The question of why the
cosmological constant strictly vanishes will still have to be
understood, either from physical ground or by invoking a
multiverse-like structure. If all the tests are negative, whatever
the precision of the observation, then the $\Lambda$CDM will
remain a cosmological model on which we will have no handle.

Constructing a cosmological model which will make sense both from
phy\-sics and the observations still require a lot of work. Many
models can save the phenomena but none are based on firm physical
grounds. The fact that we have to invoke structures on scales much
larger than those than can be probed to make sense of the
acceleration of the cosmic expansion may indicate that we may be
reaching a limit of what physical cosmology can explain.

\end{document}